\documentclass[a4paper,twocolumn,showpacs,superscriptaddress,floatfix,nofootinbib,prd]{revtex4-1}

\usepackage{graphicx}
\usepackage{epsf}
\usepackage{epsfig}
\usepackage{amssymb,amsmath}
\usepackage[usenames]{color}
\usepackage{amssymb}
\usepackage{times}
\usepackage{mathrsfs}
\usepackage{hyperref}
\hypersetup{
  colorlinks=true,        
  linkcolor=blue,         
  citecolor=cyan,         
}
\usepackage{float}

\usepackage{tensind}
\tensordelimiter{?}

\definecolor{orange}{rgb}{1,0.5,0}

\newcommand{\mn}[1]{\texttt{#1}}

\newcommand{\cf}{cf.,~}
\newcommand{\ie}{i.e.,~}
\newcommand{\eg}{e.g.,~}
\newcommand{\dd}{\mathrm{d}}
\renewcommand{\BibitemShut}[1]{}

\newcommand{\s}{\rm s}
\newcommand{\ms}{\rm ms}
\newcommand{\km}{\,{\mathrm{km}}}
\newcommand{\gcm}{\,{\mathrm{g}/\mathrm{cm}^{3}}}
\newcommand{\Msun}{M_{\odot}}
\newcommand{\Msol}{M_{\odot}}
\newcommand{\Ye}{Y_\mathrm{e}}
\newcommand{\kB}{k_{_\mathrm{B}}}
\newcommand{\Mpc}{\,{\mathrm{Mpc}}}
\graphicspath{{figs/}}
\begin{document}


\title[On r-process nucleosynthesis from binary neutron star mergers]{On
  r-process nucleosynthesis from matter ejected in binary neutron star
  mergers}

\author{Luke~Bovard}
\affiliation{Institut f{\"u}r Theoretische Physik,
Johann Wolfgang Goethe-Universit{\"a}t, Max-von-Laue-Stra{\ss}e 1, 60438 Frankfurt,
Germany}
\author{Dirk~Martin}
\affiliation{Institut f{\"u}r Kernphysik, Technische Universit{\"a}t
  Darmstadt, Schlossgartenstra{\ss}e 9, 64289 Darmstadt, Germany}
\author{Federico~Guercilena}
\affiliation{Institut f{\"u}r Theoretische Physik, Johann Wolfgang
  Goethe-Universit{\"a}t, Max-von-Laue-Stra{\ss}e 1, 60438 Frankfurt,
  Germany}
\author{Almudena~Arcones}
\affiliation{Institut f{\"u}r Kernphysik, Technische Universit{\"a}t
  Darmstadt, Schlossgartenstra{\ss}e 9, 64289 Darmstadt, Germany}
\author{Luciano~Rezzolla}
\affiliation{Institut f{\"u}r Theoretische Physik, Johann Wolfgang
  Goethe-Universit{\"a}t, Max-von-Laue-Stra{\ss}e 1, 60438 Frankfurt,
  Germany}
\affiliation{Frankfurt Institute for Advanced Studies,
  Ruth-Moufang-Stra{\ss}e 1, 60438 Frankfurt, Germany}
\author{Oleg~Korobkin}
\affiliation{Center for Theoretical Astrophysics, Los Alamos National
  Laboratory, Los Alamos, NM 87545, USA}

\begin{abstract}
When binary systems of neutron stars merge, a very small fraction of
their rest mass is ejected, either dynamically or secularly. This
material is neutron-rich and its nucleosynthesis could provide the
astrophysical site for the production of heavy elements in the universe,
together with a kilonova signal confirming neutron-star mergers as the
origin of short gamma-ray bursts. We perform full general-relativistic
simulations of binary neutron-star mergers employing three different
nuclear-physics EOSs, considering both equal- and unequal-mass
configurations, and adopting a leakage scheme to account for neutrino
radiative losses. Using a combination of techniques, we carry out an
extensive and systematic study of the hydrodynamical, thermodynamical,
and geometrical properties of the matter ejected dynamically, employing
the \texttt{WinNet} nuclear-reaction network to recover the relative
abundances of heavy elements produced by each configurations. Among the
results obtained, three are particularly important. First, we find that
both the properties of the dynamical ejecta and the nucleosynthesis
yields are robust against variations of the EOS and masses, and match
very well the observed chemical abundances. Second, using a conservative
but robust criterion for unbound matter, we find that the amount of
ejected mass is $\lesssim 10^{-3}\,M_{\odot}$, hence at least one order
of magnitude smaller than what normally assumed in modelling kilonova
signals. Finally, using a simplified and gray-opacity model we assess the
observability of the infrared kilonova emission finding, that for all
binaries the luminosity peaks around $\sim1/2$ day in the $H$-band,
reaching a maximum magnitude of $-13$, and decreasing rapidly after one
day. These rather low luminosities make the prospects for detecting
kilonovae less promising than what assumed so far.
\end{abstract}


\maketitle

\section{Introduction}
\label{sec:INTRO}

The recent detections of gravitational waves from binary black hole
mergers \cite{Abbott2016a,Abbott2016g,Abbott2017a} by LIGO has signalled
the beginning of the era of gravitational-wave astronomy. Additional
detectors such as Virgo, KAGRA and the Einstein Telescope (ET)
\cite{Accadia2011_etal, Kuroda2010, Punturo:2010} are coming online or
projected for operation in the next few years and will allow for a new
observational window on the universe, complementary to the
electromagnetic one.

An exciting possibility opened up by these advancements is the
simultaneous detection of an electromagnetic counterpart corresponding to
a gravitational wave detection from a binary neutron star (BNS) merger,
which could help explain the long-standing puzzle of the origin of short
gamma-ray bursts (SGRBs) \cite{Narayan92, Eichler89,
  Rezzolla:2011,Bartos:2012, Berger2013b}. Although only black hole
mergers have so far been detected, BNS mergers are expected to be
observed in the coming years. As such, significant progress has been made
over the last decade to accurately simulate their inspiral, merger and
post-merger dynamics (see Refs. \cite{Baiotti2016,Paschalidis2016} for
some recent reviews).

An electromagnetic counterpart from a merger that has recently received
significant attention is that of a kilonova \cite{Li1998, Rosswog2013,
  Piran2013, Grossman2014, Perego2014, Wanajo2014, Just2015,
  Sekiguchi2015, Radice2016, Just2016, Sekiguchi2016, Metzger2016,
  Tanaka2016, Barnes2016, Rosswog2017, Wollaeger2017}. A kilonova is an
infrared/optical signal powered by the decay of a variety of heavy
elements, with a dominant contributions from the elements near the second
$r$-process peak (\ie $^{133}{\rm I}$, $^{132}{\rm Te}$ and $^{133}{\rm
  Xe}$), and subdominant ones from the third $r$-process peak and
unstable transuranian elements. These elements can be formed after a BNS
merger due to the onset of rapid neutron-capture process (r-process; see
Ref.~\cite{Metzger2017} for a recent review). Kilonovae have potentially
already been observed in GRB 130603B \cite{Berger2013, Tanvir2013}, GRB
060614 \cite{Yang2015, Jin2015} and GRB 050709 \cite{Jin2016}, but the
very large uncertainties in these measurement have so far prevented an
unambiguous identification.

The power source of kilonovae is the decay of elements produced during
the r-process and throughout the history of our universe this process has
given rise to about half of the elements heavier than iron. While its
fundamental concept has been known for decades \cite{Burbidge1957}, its
astrophysical origin has not been unambiguously identified yet. For
matter to undergo r-process nucleosynthesis, in fact, a very neutron-rich
and explosive environment is required and this puts constraints on the
potential astrophysical sites where the process should take place. The
two commonly suggested astrophysical sites are core-collapse supernovae
and BNS mergers. Recent simulations of core-collapse supernovae (CCSN) have
shown that the environment in the outer layers of the explosion is not
neutron-rich enough and have been unable to reproduce the observed solar
system abundances of heavy elements \cite{Huedepol2010, Huedepol2010a,
Fischer2010, Wanajo2013}, although rare forms of CCSN driven by magnetic
fields are also a possibility \cite{Winteler2012, Moesta2014,
Nishimura2017}. In contrast, neutron star mergers are considered an
increasingly likely source of heavy elements. Recent observations of
ultrafaint dwarf galaxies \cite{Ji2016} have strongly pointed towards BNS
mergers being the main site of production of r-process elements.

Furthermore, increasingly sophisticated numerical-relativity simulations
with neutrino transport have shown that not only significant amounts of
material are ejected (due to a variety of physical processes) in BNS
mergers, but the environment in the ejecta provides the necessary
conditions to trigger and sustain robust r-process nucleosynthesis.
Numerous simulations ranging from Newtonian to full relativistic, with a
variety of microphysical treatments, have shown four broad ejection
mechanisms. These are: dynamical ejecta \cite{Rosswog1999,Rezzolla:2010,
Roberts2011, Kyutoku2012, Rosswog2013a, Bauswein2013b, Foucart2014,
Hotokezaka2013, Wanajo2014, Sekiguchi2015, Sekiguchi2016, Radice2016,
Lehner2016, Dietrich2016}, neutrino-driven winds \cite{Dessart2009,
Perego2014, Just2015,Martin2015, Martin2015a, Just2016,
Murguia-Berthier2014, Fujibayashi2017}, magnetically driven winds
\cite{Shibata2011b, Kiuchi2012b, Siegel2014, Rezzolla2014b, Ciolfi2014},
and viscous evolution of the accretion disk
\cite{Beloborodov2008,Metzger2008a,Goriely2011,Fernandez2013}. Their
typical time scales are approximately $\sim 10\,\ms$ for dynamical
ejecta, $\sim 100\,\ms$ for magnetically driven or neutrino-driven winds,
and $\sim 1\,\s$ for viscous evolution. Due to the high computational
cost of performing long-term fully relativistic simulations, mostly
dynamical ejecta have been studied in full relativity, while other
mechanisms have been the subject of mostly Newtonian simulations.

In this work, we present a number of high-resolution numerical-relativity
simulations of BNS mergers to investigate the effects of the neutron-star
initial masses, mass ratios and most importantly the microphysical
equation of state (EOS) on the resulting r-process nucleosynthesis. We
consider three fully temperature-dependent EOSs spanning a wide range of
stiffness. For each EOS, we consider three equal-mass initial setups
covering a realistic range of initial BNS masses. Additionally, we
consider for each EOS one unequal-mass case.

To follow the evolution of the fluid, we use a combination of techniques,
namely outflow detectors and passively advected fluid tracers. The
properties and use of the latter in general-relativistic simulations have
been discussed in Ref.~\cite{Bovard2016}. We then post-process the data
using a complete nuclear-reaction network
\cite{Winteler2012,Korobkin2012} to obtain the final r-process
abundances. We also compute the associated kilonova light curves using
the model outlined in Ref.~\cite{Grossman2014}.

We find that the amount of dynamically ejected mass is of the order of
$10^{-3}\,\Msun$, which, although rather small, is consistent with
current constraints on the typical BNS merger rates and observed
abundances of heavy elements in the Milky Way. Although some variation in
the properties of the ejected mass (\ie typical values of the electron
fraction, entropy or velocity) are observed and appear to correlate with
the choice of EOS or neutron-star mass for a given BNS model, these
differences have minimal influence on the final r-process nucleosynthesis
yields. Given the kilonova light curves associated to our simulations, we
find that the prospects for their direct observation are rather limited;
however, in view of the approximations made in our current analysis, this
may be not a conclusive statement. Finally, we have uncovered an
interesting geometrical structure in the angular distribution of the
ejecta which could have important implications on the properties of the
kilonova signal.

The paper is structured as follows: in Sec.~\ref{sec:ID}, we introduce
the mathematical and numerical methods employed, together with the
initial BNS configurations that we evolve. Section~\ref{sec:METHODS},
instead, summarizes the main properties of the physical models and
numerical techniques that we employ to study the BNS evolution as well as
to recover the heavy-element abundances. Sections
\ref{sec:overview}--\ref{sec:velocity} present our results and findings
in terms of the mass ejected, the electron fraction, the specific
entropy, and the ejecta velocity. Similarly,
Secs.~\ref{sec:kilonova}--\ref{sec:rates} report our estimates for the
kilonova light curves and their detectability, together with the
constraints on the merger rates of BNSs. Finally, we conclude in
Sec.~\ref{sec:CONCLUSIONS}.

Unless otherwise specified, we use a system of units such that $c = G =
\Msol = 1$, where $c$ is the speed of light in vacuum, $G$ is the
gravitational constant, and $\Msol$ is the mass of the Sun. We use
Einstein's convention of summation over repeated indices. Latin indices
run over $1,2,3$, while Greek indices run over $0,1,2,3$. The spacetime
metric signature we adopt is $(-,+,+,+)$.

\section{Physical setup and initial data}
\label{sec:ID}

\begin{table*}[t]
  \begin{tabular}{|l||cccccccccccccc|}
    \hline
    & EOS & $q$ & $M_{1}$ &  $M_{2}$ & $R_{1}$ & $R_{2}$ & $M_{_{\mathrm{ADM}}}$ &
    $M_{\rm b, 1}$  & $M_{\rm b, 2}$ & $M_{_{\mathrm{TOV}}}$ & $R_{_{\mathrm{TOV}}}$ & $\mathcal{C}_{1}$ & $\mathcal{C}_{2}$ & $J$\\
    Model & & & $[\Msun]$ &  $[\Msun]$ & $[\km]$  & $[\km]$ &
    $[\Msun]$& $[\Msun]$ & $[\Msun]$ & $[\Msun]$ & $[\km]$ &- & - & $[\Msun^2]$\\
    \hline
    \mn{LS220-M1.25} & LS220 & $1.0$ & $1.25$ & $1.25$ & $12.80$ & $12.80$ & $2.48$ & $1.36$ & $1.36$ & $2.04$ & $10.65$ & $0.144$ & $0.144$ & $6.42$\\
    \mn{LS220-M1.35} & LS220 & $1.0$ & $1.35$ & $1.35$ & $12.75$ & $12.75$ & $2.67$ & $1.47$ & $1.47$ & $2.04$ & $10.65$ & $0.156$ & $0.156$ & $7.26$\\
    \mn{LS220-M1.45} & LS220 & $1.0$ & $1.45$ & $1.45$ & $12.67$ & $12.67$ & $2.87$ & $1.60$ & $1.60$ & $2.04$ & $10.65$ & $0.169$ & $0.169$ & $8.20$\\
    \mn{LS220-q09}   & LS220 & $0.9$ & $1.21$ & $1.35$ & $12.81$ & $12.75$ & $2.61$ & $1.32$ & $1.47$ & $2.04$ & $10.65$ & $0.140$ & $0.156$ & $6.98$ \\
    \hline
    \mn{DD2-M1.25} & DD2 & $1.0$ & $1.25$ & $1.25$ & $13.20$ & $13.20$ & $2.48$ & $1.35$ & $1.35$ & $2.42$ & $11.90$ & $0.140$ & $0.140$ & $6.40$\\
    \mn{DD2-M1.35} & DD2 & $1.0$ & $1.35$ & $1.35$ & $13.23$ & $13.23$ & $2.68$ & $1.47$ & $1.47$ & $2.42$ & $11.90$ & $0.151$ & $0.151$ & $7.31$\\
    \mn{DD2-M1.45} & DD2 & $1.0$ & $1.45$ & $1.45$ & $13.25$ & $13.25$ & $2.87$ & $1.59$ & $1.59$ & $2.42$ & $11.90$ & $0.161$ & $0.161$ & $8.19$\\
    \mn{DD2-q09}   & DD2 & $0.9$ & $1.22$ & $1.35$ & $13.19$ & $13.23$ & $2.55$ & $1.31$ & $1.47$ & $2.42$ & $11.90$ & $0.136$ & $0.151$ & $6.68$ \\

    \hline
    \mn{SFHO-M1.25} & SFHO & $1.0$ & $1.25$ & $1.25$ & $11.97$ & $11.97$ & $2.48$ & $1.36$ & $1.36$ & $2.06$ & $10.31 $& $0.155$ & $0.155$ & $6.40$\\
    \mn{SFHO-M1.35} & SFHO & $1.0$ & $1.35$ & $1.35$ & $11.92$ & $11.92$ & $2.68$ & $1.48$ & $1.48$ & $2.06$ & $10.31 $& $0.167$ & $0.167$ & $7.28$\\
    \mn{SFHO-M1.45} & SFHO & $1.0$ & $1.45$ & $1.45$ & $11.87$ & $11.87$ & $2.87$ & $1.61$ & $1.61$ & $2.06$ & $10.31 $& $0.181$ & $0.181$ & $8.20$\\
    \mn{SFHO-q09}   & SFHO & $0.9$ & $1.22$ & $1.35$ & $11.97$ & $11.92$ & $2.55$ & $1.32$ & $1.48$ & $2.06$ & $10.31 $& $0.150$ & $0.167$ & $6.67$ \\

    \hline
  \end{tabular}
  \caption{Summary of the properties of the systems under consideration.
The columns denote, respectively: the EOS; the gravitational mass ratio
$q:= M_{1}/M_{2}$ at infinite separation; the gravitational masses
$M_{1,2}$ of the two stars at infinite separation; the stars' radii
$R_{1,2}$ at infinite separation; the ADM mass $M_{_{\mathrm{ADM}}}$ of
the system; the baryon masses $M_{\mathrm{b},1,2}$; the maximum mass of a
non-rotating model of the given EOS $M_{_{\mathrm{TOV}}}$; the radius of
the maximum mass non-rotating model of the given EOS
$R_{_{\mathrm{TOV}}}$; the compactnesses ${\mathcal C_{1,2}}:=
M_{1,2}/R_{1,2}$; the total angular momentum $J$ at the initial
separation. }
    \label{tab:models}
\end{table*}

We consider both equal- and unequal-mass BNS systems on quasi-circular
orbits, with initial configurations constructed from three different
EOSs, spanning a wide range in stiffness. From the stiffest to the
softest, these EOSs are: (i) DD2 \cite{Typel2010}; (ii) LS220
\cite{Lattimer91} with compressibility parameter $K = 220\,{\rm MeV}$;
SFHO \cite{Steiner2013}. Note that recent calculations in
Ref. \cite{Kolomeitsev2016} have shown that the LS220 EOS does not
satisfy constraints stemming from a lower bound on the energy per nucleon
provided by the unitary-gas approximation. This result disfavours the
LS220 as a viable model for the microphysics of neutron stars, but since
this EOS is also one of the most well-studied in numerical applications,
we include it in our study since it provides a useful comparison with the
literature. Additionally, the DD2 and SFHO EOSs include additional light
nuclei that are not included in the LS220 and these change the neutrino
interactions \cite{stellarcollapse}.

For each EOS, we consider three different equal-mass setups, with
neutron-star gravitational masses of $1.25,1.35$ and $1.45\,\Msun$,
respectively; and one unequal-mass system, with star masses of $1.2$ and
$1.35\,\Msun$, resulting in a mass ratio $q=0.9$ and a total ADM mass
(see Ref.~\cite{Gourgoulhon2012} for a definition) of the system which is
intermediate between the two lightest equal-mass configurations for the
same EOS. The stars' initial separation is chosen to be $45\km$,
resulting in an inspiral phase of approximately $\sim3$
orbits. Table~\ref{tab:models} summarizes the properties of each
system. The stars initial states are computed at neutrinoless beta
equilibrium, \ie at zero neutrino chemical potential, thus setting the
initial values of the electron fraction. The initial data for every
binary was constructed using the \verb+LORENE+ pseudo-spectral elliptic
solver \cite{Gourgoulhon01} and refers to irrotational binaries in
quasi-circular orbit.

\section{Methods}
\label{sec:METHODS}

We summarize in this section the salient features of the physical models
we employ to study the evolution of the BNS systems introduced in the
previous section, as well as the numerical methods used and their
implementation. In the interest of brevity, and since our approach does
not significantly differ from well-known ones already described in the
literature, we provide here only a succinct discussion.

\subsection{General-relativistic hydrodynamics and neutrino transport}
\label{sec:hydro}

We model the neutron-star matter (as well as the matter ejected by the
system) as a perfect fluid, using the temperature-dependent EOSs
mentioned in the previous section. The fluid evolution is described by
the continuity equation, which expresses the conservation of baryon mass,
and the relativistic Euler equations (taking the form of local
conservation of the fluid stress-energy tensor components) (see
Ref.\cite{Rezzolla_book:2013} for a comprehensive discussion).

We include the contribution from neutrino interactions, which can change
the composition of the material, and in particular the value of the
electron fraction, which would be otherwise simply advected by the fluid
velocity. To this end we employ a ``leakage'' scheme \cite{vanRiper1981,
  Ruffert96b,Rosswog:2003b}, which takes into account cooling due to
neutrino emission, but does not model absorption and heating. In the
presence of such interactions, a source term must be added both to the
continuity equation and Euler equations, which, following
\cite{Galeazzi2013}, take the form
\begin{align}
   \nabla_\alpha (n_{\rm b}\, u^\alpha) &= 0\,, \\
   \nabla_\alpha (n_{\rm e}\, u^\alpha) &= R\,, \\
   \nabla_\beta T^{\alpha\beta}     &= Qu^{\alpha}\,,
   \label{eq:continuityandeuler}
\end{align}
where $n_{\rm b}$ and $n_{\rm e}$ are the baryon and electron number
density, $u^\alpha$ is the fluid 4-velocity and $T^{\alpha\beta}$ is the
fluid stress-energy tensor. Here, $R$ is the net lepton-number emission
rate, while $Q$ is the net neutrino-cooling rate, and both are defined
per unit volume and in the fluid rest-frame. A detailed discussion on the
estimation of $Q$ and $R$ is contained in Refs.~\cite{Galeazzi2013,
  Radice2016}.

The numerical scheme used to solve the hydrodynamics evolution equations
is a finite-volumes method, applied to the flux-conservative formulation
of Eqs.~\eqref{eq:continuityandeuler}. We employ the fifth-order MP5
\cite{suresh_1997_amp} reconstruction operator, the HLLE Riemann solver
\cite{Harten83} and the positivity-preserving limiter of
Ref.~\cite{Hu2013,Radice2013c}. We also make use of the refluxing technique
\cite{Berger89} to minimize numerical spurious losses or gains of mass at
the interface between refinement levels. The scheme is implemented in the
\texttt{WhiskyTHC} code \cite{Radice2012a,Radice2013c}.

To integrate Einstein equations and obtain the spacetime evolution we use
a fourth order finite-differences method applied to the BSSNOK
formulation \cite{Shibata95,Baumgarte99,Brown09} of Einstein equations.
The gauge conditions are the standard ``1+log'' and ``Gamma driver''
choices (see, \eg Ref.~\cite{Baumgarte2010}). The spacetime evolution is
provided by the \texttt{Mclachlan} code \cite{Brown:2008sb}, and coupled
to the hydrodynamics evolution through the evaluation of the fluid
stress-energy tensor.

An adaptive mesh refinement (AMR) approach based on the \texttt{Carpet}
mesh-refinement driver~\cite{Schnetter-etal-03b} is used to increase
resolution as well as extend the spatial domain, placing the outer
boundary as close as possible to the wave zone. In particular, we employ
a Cartesian 3D grid with six box-in-box levels of mesh refinement
(promoted to seven after merger), so that the finest, innermost grid
during the inspiral has a resolution of $0.15\, \Msun \simeq 215 \, {\rm
m}$. The outer boundary of the domain extends to $512\, \Msun \simeq
760\, \km$. The timestep is fixed to one sixth of the grid spacing and a
third-order strong stability preserving Runge-Kutta method is used for
advancing the computation in time.

\subsection{Tracer particles and outflow detectors}
\label{sec:traceroutflow}

To follow the flow of ejected material we employ two different
techniques. The first technique is the use of tracer particles
\cite{Wanajo2014,Kastaun2016,Mewes2016,Bovard2016}, \ie massless
particles passively advected with the fluid. A total of $2\cdot 10^5$
tracers are placed with a uniform distribution in the density interval
$10^{7}\, \mathrm{g/cm}^3\lesssim\rho\lesssim 10^{15}\, \mathrm{g/cm}^3$
at the time of merger (see Ref.~\cite{Bovard2016} for a discussion on why
this distribution of tracers is the optimal one). Fluid properties are
interpolated at the tracers location, providing a detailed account of the
evolution of the associated fluid element. Following the description in
Ref.~\cite{Bovard2016}, a ``tracer mass'' can be associated to the
otherwise massless tracers by locally integrating a mass flux through a
sphere of given radius. Combining this mass with the history of the
evolution of the tracer particle provides the initial input for the
nuclear-reaction network discussed in Sec. \ref{sec:nuclear-network}.

The second technique employed to follow the ejected material is the use
of so-called outflow detectors, \ie spherical surfaces placed at a fixed
coordinate radius around the center of the computational domain. These
detectors are able to measure the flux of the fluid through their surface
and record the various hydrodynamical and thermodynamical quantities as a
function of time. In our simulations, we employ nine detectors set at
radii between $100$ and $500\,\Msun$ with a separation of $50\,\Msun$.
Each detector has a resolution of 55 points in the polar and 96 points in
the azimuthal direction, and the detector located at a radius of $200
\Msun\approx 300\km$ is our fiducial one. As the fluid passes through a
detector spherical surface, hydrodynamical and thermodynamical variables
are interpolated onto it, allowing us to record the entire evolution of
the fluid in all angular directions. Note that we define the total
ejected mass by integrating the unbound mass flux over the surface of the
detector, in contrast to, \eg Ref.~\cite{Hotokezaka2013}, where the
rest-mass density of all unbound fluid elements is integrated over the
whole computational domain (see Sec.~\ref{sec:mass-eject} for further
details).

\subsection{Selection of unbound material}
\label{sec:unboundcrit}

Regardless of whether tracer particles or outflow detectors are used, it
is necessary to define a criterion to identify gravitationally unbound
material, which will not accrete back onto the merger remnant and can be
considered ejected from the system.

The difficulty in determining gravitationally unbound material arises
mostly due to the finite size of the grid. Ejecta can only be followed to
the edge of the computational domain, which is still relatively close to
the BNS merger product, and can still be influenced by its gravitational
potential. This problem could be alleviated by using a larger grid, but
this comes at greater computational cost and a few numerical drawbacks
(\eg poor resolution in an AMR grid). Likewise, we are interested in
tracking the evolution of the ejected material to study the kilonova
signal, which is expected to peak days after merger. However, computing
the evolution of the ejecta for such long timescales is currently
computationally unfeasible in full numerical-relativity simulations,
which can run at most for timescales on the order of tens of milliseconds
after merger.

As such, a criterion to define unbound material is therefore needed and
we choose to define a fluid element as ``unbound'' if it satisfies the
so-called \textit{geodesic criterion} (\eg Refs.~
\cite{Kastaun2014,Sekiguchi2015}), \ie if $u_{t}\le -1$, where $u_{t}$ is
the covariant time component of the fluid element 4-velocity. The
justification of such a criterion is clear when considering its Newtonian
limit. In this case $u_{t} \approx -1 - \phi - v^{2}/2$, where $\phi$ is
the gravitational potential (see Ref.~\cite{Rezzolla_book:2013}). At
large separations from the gravitational sources, the gravitational
potential can be neglected, $\phi \simeq 0$ and thus $u_{t} \approx -1
-v^{2}/2\le -1$. The criterion amounts therefore to imposing that the
fluid element should have non-zero velocity at infinity.

An alternative criterion that has been studied \cite{Kastaun2014} is the
so-called \textit{Bernoulli criterion}. In this case, a fluid element is
defined to be unbound if $hu_{t}\le -1$, $h$ being the fluid specific
enthalpy. In the following, we only consider the geodesic criterion, and
hereafter the adjective ``unbound'' will refer exclusively to material
satisfying it. We consider however the impact that the choice of the
criterion for material to be unbound can have on the properties of dynamically
ejected material in Appendix \ref{sec:bernoulli}, where we present a
comparison of the results obtained with the geodesic and Bernoulli
criteria. However, since $h \geq 1$ \cite{Rezzolla_book:2013}, it is
clear that the Bernoulli criterion will be in general less restrictive
than the geodesic one, yielding an amount of ejected material that is at
least twice larger.

\subsection{Nuclear network overview}
\label{sec:nuclear-network}

The nucleosynthesis calculations are carried out with the complete
\texttt{WinNet} nuclear-reaction network \cite{Winteler2012,
  Korobkin2012}. Over 5800 nuclei between the valley of stability and the
neutron-drip line are taken into account. The reaction rates are taken
from the compilation of Ref.~\cite{Rauscher2000} for the Finite Range
Droplet Model (FRDM \cite{Moeller1995}) and we consider weak-interaction
rates including neutrino absorption on nucleons \cite{Moeller2003,
  Froehlich2006}. Neutron-capture rates for nuclei with atomic number
$Z\gtrsim 80$ and neutron-induced fission rates are taken from
Ref.~\cite{Panov2010}. Moreover, we include beta-delayed fission
probabilities from Ref.~\cite{Panov2005}. Our network has been used as a
benchmark in a recent comparison with another general-purpose
nuclear-reaction network \cite{Lippuner2017}, showing a very good overall
agreement.

As detailed in Sec. \ref{sec:abundances}, we post-process representative
subsets of unbound tracers from the hydrodynamical simulations according
to three different methods of selection. From every tracer, a time series
of the rest-mass density, temperature, specific entropy, and electron
fraction is extracted, on which the nuclear network acts. For each of
these tracers, we start our calculations when the temperature drops below
$T = 10^{10}\,{\rm K} = 10\,\mathrm{GK}$. Due to the high temperatures, the
initial composition is given by nuclear statistical equilibrium (NSE),
and is dominated by nucleons and alpha particles. We assume NSE to hold
for $T \gtrsim 8~\mathrm{GK}$. When the temperature drops below the NSE
threshold, the composition is evolved with the full reaction network. As
most of the tracer trajectories were simulated only until $\sim
20~\mathrm{ms}$ after the merger, we extrapolate them to very large
distances using the following prescriptions for the position, density and
temperature evolution \cite{Fujimoto2008, Korobkin2012}
\begin{align}
    r(t) &= r_0 + v_0t,
    \label{eq:expansion-r}\\
    \rho(t) &= \rho_0 \left(\frac{t}{t_0}\right)^{-3},
    \label{eq:expansion-hor}\\
    T(t) &= T\left[s,\rho(t),\Ye(t)\right]\,.
    \label{eq:expansion-T}
\end{align}
where $\rho$ is the total rest-mass density, $r$ the coordinate radius,
$v$ the 3-velocity, $s$ the specific entropy, and $\Ye:=n_{\rm e}/n_{\rm
  b}$ the electron fraction. The subscript ``$0$'' indicates the last
available values from the hydrodynamical simulations, and the temperature
is computed from the Helmholtz EOS \cite{Timmes1999, Timmes2000}. This
ansatz for the ejecta expansion is well justified, at least at late
times, as shown in Ref.~\cite{Bovard2016}, where tracers were reported to
move ballistically along radial directions and to expand adiabatically at
large distances from the merger product.

Furthermore, we compute the energy released by the r-process and include
its impact on the evolution of the fluid entropy
\cite{Freiburghaus:1999}. In particular, the major contribution to the
radioactive heating is expected to come from beta decays and we assume
the energy to be about equally distributed between thermalising electrons
and photons, and escaping neutrinos and photons \cite{Metzger:2010}.

\section{Overview of simulations}
\label{sec:overview}

In what follows we discuss the results from the simulations comparing the
outflow properties of the dynamically ejected material such as: the mass
ejected $M_{\rm ej}$, the electron fraction $\Ye$, the specific entropy
$s$, and the ejecta velocity $v_{\rm ej}$, for the different simulation
parameters.

To investigate the effects of the EOS and initial masses on the dynamical
ejecta, and hence the r-process nucleosynthesis, a total of 12
simulations were run. To study the effects of the EOS, three fully
temperature-dependent EOSs were used, spanning a wide range in stiffness.
For each EOS, four different masses parameters were run with $3$
equal-mass and $1$ unequal-mass case. For each simulation, at least
$10\,\ms$ after merger was simulated to ensure a sufficient time for the
dynamical ejecta to reach $300\km$, which is where the properties of the
dynamical ejecta are measured.

For each BNS model, we simulate approximately $\sim3$ orbits before
merger and we define the time of merger to be the time at which the
gravitational-wave amplitude reaches its first peak \cite{Baiotti:2010};
in the following we define the time origin such that $t=0$ corresponds to
the time of merger. Given the maximum mass of non-rotating neutron star
models, $M_{_{\rm TOV}}$, and the initial mass of the merging binaries,
all the mergers that do not yield a prompt collapse to a black hole
produce a hypermassive neutron star (HMNS), \ie a neutron star whose mass
exceeds the maximum mass supported by uniform rotation, $M_{\rm max}
\simeq 1.20 M_{_{\rm TOV}}$ \cite{Breu2016}, and that is in a metastable
equilibrium state supported by differential rotation, with a
quasi-universal rotation profile \cite{Hanauske2016}.

The three binaries which instead collapse to a black hole are
\texttt{SFHO-M1.35}, \texttt{SFHO-M1.45}, and \texttt{LS220-M1.45} with
the latter two being a prompt collapse.  More specifically, for
\texttt{SFHO-M1.45} the collapse is right at merger and results in very
little material being ejected (see discussion in
Sec.~\ref{sec:mass-eject}), while for \texttt{LS220-M1.45} the collapse
takes place about $\sim0.5\,\ms$ after merger, which is sufficient to
allow for material to be ejected. Finally for \texttt{SFHO-M1.35}, the
collapses to a black hole takes place at $\sim 10\,\ms$ after the merger,
when the HMNS has lost sufficient angular momentum.

\begin{figure*}
\begin{center}
\includegraphics[width=2.0\columnwidth]{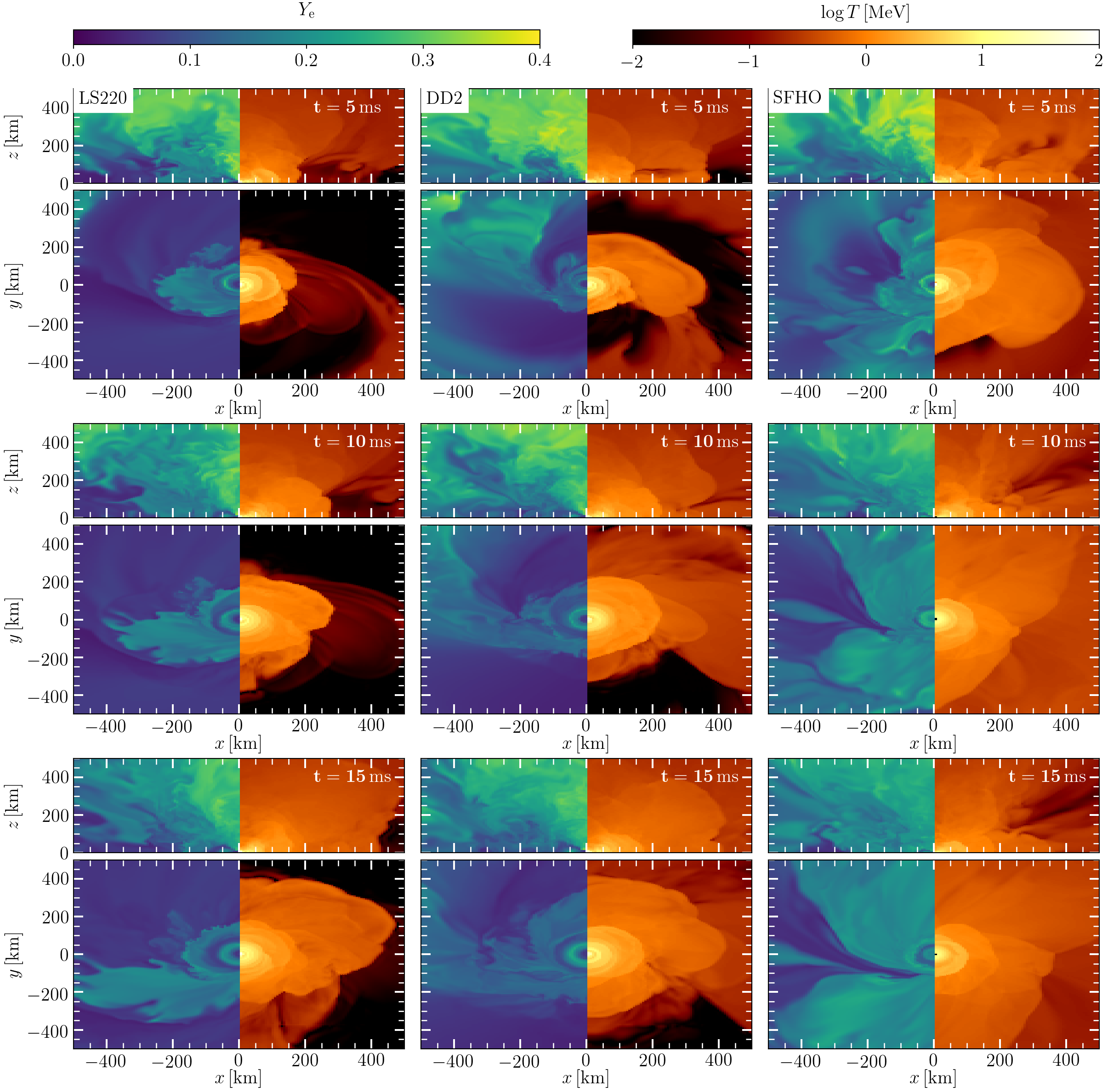}
\caption{Evolution of the electron fraction (left parts of panels) and of
  the temperature (right parts of panels) on the $(x,z)$ plane (top
  panels) and on the $(x,y)$ plane (bottom panels), for the different
  EOSs, namely: LS220, DD2, SFHO, from left to right. All panels refer to
  binaries with masses of $2\times1.35\Msol$ and at the same
  representative times: $5\,\ms$ (top row), $10\,\ms$ (middle row), and
  $15\,\ms$ (bottom row) after the merger.}
\label{fig:ye-temp}
\end{center}
\end{figure*}

To show the spatial distributions of various quantities in the
simulations, Fig.~\ref{fig:ye-temp} reports three different time slices,
$5,10,15\,\ms$ from top to bottom, of the electron fraction (left panels)
and the temperature (right panels) in the $(x,y)$- (bottom panels) and
$(x,z)$-planes (top panels) for the three different EOSs of $1.35\Msol$
equal-mass initial data.

\begin{table*}
\begin{tabular}{|l||cccccccccc|}
\hline
&  $M_{\rm{ej}}$ & $\langle \Ye \rangle $ & $\langle s
\rangle $ & $\langle v_{\rm{ej}} \rangle $ &
$\langle v_{\infty} \rangle $ &  $t_{H,{\rm peak}}$ & $L_{\rm peak}$ &
$m_{J,{\rm peak}}$ & $m_{H,{\rm peak}}$ & $m_{K,{\rm peak}}$ \\
Model & $[10^{-3}\, \Msol]$& -  & $[\kB]$ & $[10^{-1} \rm{c}]$ & $[10^{-1} \rm{c}]$ & $[\rm{days}]$ &
$[10^{40}\, \rm{erg/s}]$ & $[{\rm{AB}}]$ & $[{\rm{AB}}]$ & $[{\rm{AB}}]$ \\
\hline
\mn{LS220-M1.25}  & $0.61$ & $0.08$ & $10.3$ & $2.2$ & $1.6$ & $0.53$ & $2.24$ & $-12.6$ & $-12.6$ & $-12.4$ \\
\mn{LS220-M1.35}  & $0.82$ & $0.10$ & $12.7$ & $2.2$ & $1.5$ & $0.51$ & $2.00$ & $-12.5$ & $-12.4$ & $-12.2$ \\
\mn{LS220-M1.45}  & $1.09$ & $0.11$ & $10.5$ & $2.6$ & $2.1$ & $0.48$ & $2.62$ & $-12.8$ & $-12.7$ & $-12.5$ \\
\mn{LS220-q09}    & $0.90$ & $0.09$ & $11.9$ & $2.2$ & $1.5$ & $0.50$ & $1.94$ & $-12.4$ & $-12.3$ & $-12.1$ \\
\hline
\mn{DD2-M1.25}    & $0.96$ & $0.13$ & $13.9$ & $2.3$ & $1.7$ & $0.50$ & $2.24$ & $-12.6$ & $-12.5$ & $-12.4$ \\
\mn{DD2-M1.35}    & $0.58$ & $0.14$ & $16.5$ & $2.4$ & $1.8$ & $0.50$ & $2.44$ & $-12.7$ & $-12.7$ & $-12.5$ \\
\mn{DD2-M1.45}    & $0.50$ & $0.17$ & $19.2$ & $2.7$ & $2.1$ & $0.50$ & $2.89$ & $-12.9$ & $-12.9$ & $-12.5$ \\
\mn{DD2-q09}      & $0.46$ & $0.14$ & $18.5$ & $2.3$ & $1.7$ & $0.53$ & $2.34$ & $-12.7$ & $-12.6$ & $-12.4$ \\
\hline
\mn{SFHO-M1.25}   & $0.55$ & $0.14$ & $15.6$ & $2.5$ & $2.0$ & $0.47$ & $2.54$ & $-12.8$ & $-12.7$ & $-12.5$ \\
\mn{SFHO-M1.35}   & $3.53$ & $0.16$ & $12.7$ & $2.7$ & $2.2$ & $0.53$ & $3.36$ & $-13.2$ & $-13.2$ & $-13.0$ \\
\mn{SFHO-M1.45}   & $0.01$ & $0.24$ & $35.9$ & $3.1$ & $2.6$ & $0.16$ & $0.86$ & $-11.1$ & $-10.9$ & $-10.5$ \\
\mn{SFHO-q09}     & $0.76$ & $0.16$ & $18.8$ & $2.4$ & $1.8$ & $0.60$ & $2.92$ & $-12.0$ & $-13.0$ & $-12.9$ \\
\hline
\end{tabular}
\caption{Summary of the mass-averaged quantities of
  Sec.~\ref{sec:outflow-prop} and kilonova observational quantities of
  Sec.~\ref{sec:kilonova} computed from the simulations. The columns are,
  respectively: $M_{\rm ej}$ the dynamical mass ejecta measured at
  $300\km$, $\langle \Ye\rangle$ the mass-averaged electron fraction,
  $\langle s\rangle$ the mass-averaged entropy, $v_{\rm ej}$ the
  mass-averaged velocity of the ejecta, $\langle v_{\infty}\rangle$ the
  velocity of the ejecta at infinity using Eq.~\eqref{eq:v_inf},
  $t_{H,\rm{peak}}$ the peak time in the $H$-band of the kilonova signal,
  $L_{\rm peak}$ the peak luminosity of the kilonova, $m_{X,\rm{peak}}$
  the peak absolute magnitude in the $X=J,H,K$ bands respectively. }
\label{table:results}
\end{table*}

As anticipated in the Introduction, in terms of dynamical ejecta, there
are two main processes which can eject material: tidal forces and
shock heating. Tidal forces arise from tidal interactions during
merging and eject material primarily along the orbital plane and are
a manifestation of gravitational interactions. In comparison, shock
heating, is approximately spherically symmetric \cite{Sekiguchi2015}
and depends on the thermal properties of the fluid. These two
distinct mechanisms are illustrated in Fig.~\ref{fig:ye-temp} where
the planar region shows lower $\Ye$ and denser material, while the polar
regions have higher $\Ye$ and are more rarefied.

We first consider the tidal ejecta. This kind of ejecta tends to be very
neutron-rich, since it becomes unbound immediately during and following
merger, and originates from matter near the surfaces of the stars. These
tidal tails can be observed in the $(x,y)$-plane at $5\,\ms$ (top row)
panels of Fig.~\ref{fig:ye-temp}, where the they are visible in
the outer regions beyond $300\km$. This ejected material also tends to be
cooler, with a temperature of around $1$~MeV. In contrast, in the
$(x,z)$-plane, the $\Ye$ reaches much higher values, approximately $0.3$,
that are not observed in the orbital plane. These higher values in the
electron fraction are due to the shocked-heated material. In the polar
regions right above the HMNS, no material is ejected tidally and
neutrinos become free streaming very close to the merger product. As a
result of weak interactions by means of which the free neutrons are
converted into protons, the material becomes less neutron-rich. However,
as the angle from the pole decreases, the material becomes more optically
thick and more neutron-rich as the neutrino interactions are not as
strong. This angular dependence is also seen in the temperature profiles
as there are higher temperatures near the polar axis when compared with
the orbital plane.

It is important to note here that although neutrinos are only treated
simplistically \cite{Galeazzi2013} this broad-brush description is
qualitatively similar to more sophisticated approaches such as those
using an M1-scheme which lead to an increase in the amount of ejected
material in the polar regions \cite{Foucart2015, Foucart2015b,
Foucart2016a, Sekiguchi2015, Sekiguchi2016} and higher $\Ye$.

Turning to the effects of the EOS, there is a clear overall trend to be
deduced from Fig.~\ref{fig:ye-temp}. The ``softer'' an EOS is, the hotter
the matter tends to be. This is due to the fact that a softer EOS allows
for a deeper gravitational well, which, in turn, allows for the material
to become hotter. This dependence is clearest when comparing the softer
SFHO and the stiffer LS220 EOSs (left and right columns)\footnote{As
discussed in Sec.~\ref{sec:ID}, the inclusion of light nuclei changes the
composition, but not the temperature.}, where the temperature in the
$(x,y)$-plane is much hotter for the SFHO than the LS220, as expected. As
a result, because neutrino interactions depend on the temperature, the
electron fraction is also higher the softer an EOS is. Again this is most
clear when examining the fluid properties on the $(x,y)$-plane of the
SFHO and LS220 simulations, where the data referring to the LS220 EOS is
much more neutron-rich when compared with the SFHO.

\section{Matter-outflow properties}
\label{sec:outflow-prop}

This section is dedicated to a comprehensive discussion of the properties
of the matter that is ejected dynamically in the merger and is unbound.
In particular, we will concentrate on the total amount of ejected matter
as well as on the distributions of this matter in terms of the electron
fraction, of the specific entropy and of the velocity of the fluid
elements.

\subsection{Ejected-mass}
\label{sec:mass-eject}

\begin{figure}[!t]
\begin{center}
\includegraphics[width=1.0\columnwidth]{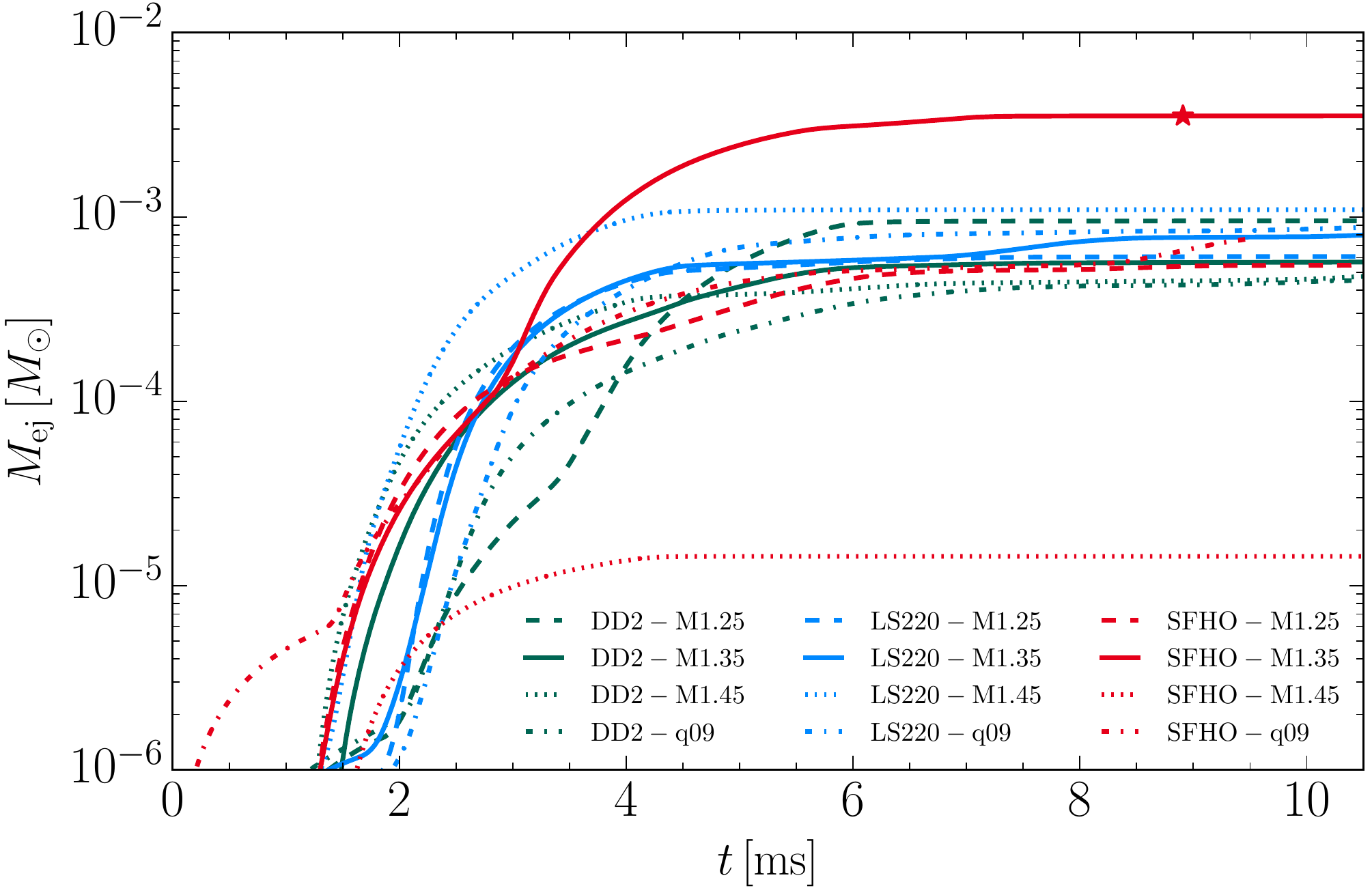}
\caption{Evolution of the dynamically ejected unbound mass $M_{\rm ej}$
  as measured through a detector at radius $300\km$ when using the
  geodesic criterion and for the various binaries considered. The star
  denotes the time of black-hole formation for model \texttt{SFHO-M1.35}.
  Binaries \texttt{LS220-M1.45} and \texttt{SFHO-M1.35} collapse shortly
  after merger and are not visible in the plot.}
    \label{fig:mass-eject}
\end{center}
\end{figure}

An accurate measure of the total amount of ejected material from a binary
merger is essential for the characterization of r-process elements and on
the potential observable properties of kilonova (see
Sec.~\ref{sec:kilonova}). In Sec. \ref{sec:traceroutflow}, we defined the
total ejected mass using outflow detectors which measure the flux of
unbound material at a given radius. Hereafter, we consider the detector
placed at a radius of $200\Msun\approx 300\km$ as the fiducial detector
through which to measure the amount and properties of the ejected
material. To compute the total mass ejected, the flux of the rest-mass
density through the detector's spherical surface is computed and then
integrated over the whole sphere. This gives the total mass-flux which
can be integrated over time to provide a measurement of the total
dynamically ejected material $M_{\rm ej}$. In this calculation, only the
flux associated to unbound fluid elements contributes to the integral.
Explicitly, for a detector at a given radial distance, the total ejected
mass is given by
\begin{equation}
    M_{\rm ej}(t) := \int_0^{t} \int_\Omega \rho_\ast
    W (\alpha v^r - \beta^r) \sqrt{\gamma}_{\Omega} \dd \Omega \dd t'\,,
    \label{eq:m_ej}
\end{equation}
where $\sqrt{\gamma}_{\Omega}$ is the surface element on the detector
(\ie the square root of the 2-metric induced on the detector by the
spacetime 4-metric); the term $\rho_\ast W (\alpha v^r - \beta^r)$ is the
flux of mass through the sphere, expressed in terms of the 3+1
quantities: the lapse function $\alpha$, the shift vector $\beta^i$, and
the fluid 3-velocity $v^i$, the Lorentz factor $W:=(1-v^i v_i)^{-1/2}$
and the fraction of the rest-mass density that is unbound $\rho_\ast$,
\ie of fluid elements that do satisfy the geodesic criterion. The
integral of the mass flux can then be integrated in time beginning at
merger, \ie $t=0$ and ending at $T_f$, the time at the end of the
simulation.

Figure~\ref{fig:mass-eject} reports the amount of ejected material
computed through Eq.~\eqref{eq:m_ej} for the LS220 (blue), DD2 (green),
and SFHO (red) EOSs and the different masses and mass
ratios\footnote{Unless specified otherwise, hereafter we will use the
same colour scheme to refer to the various EOSs: simulations with the
LS220 EOS are shown blue, DD2 in green, and SFHO in red. Furthermore,
the different masses are defined as follows, $1.25\Msol$ is dashed,
$1.35\Msol$ is solid, $1.45\Msol$ is dotted, and $q=0.9$ is
dash-dotted.}. The results of Fig.~\ref{fig:mass-eject} are also
summarised in Table~\ref{table:results}, where $M_{\rm ej}$ refers to the
mass ejected $t=10\,\ms$ after merger.

Overall, the qualitative behaviour of all simulations is similar. There is
a large ejection of material, due to tidal interactions and shock
heating, that reach the detector approximately $1\,\ms$ after merger and
continues for about $4-5\,\ms$ before the flux becomes zero. However,
this apparent decrease in ejected material is simply due to the geodesic
criterion not being satisfied by the outflowing material and not a
physical decrease in outflow. In Appendix \ref{sec:bernoulli}, we discuss
how this picture changes when considering the Bernoulli criterion, which
allows for a longer period of ejected material.

\begin{figure*}
\begin{center}
\includegraphics[width=2.0\columnwidth]{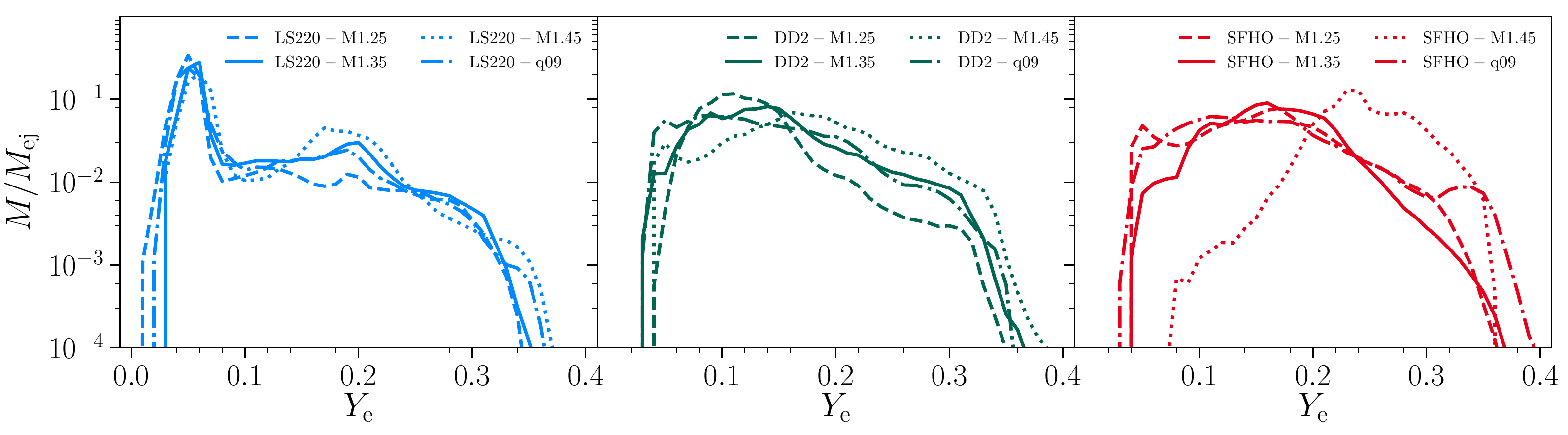}
\caption{Distributions of the ejected mass fraction as function of the
  electron fraction $\Ye$, as measured by a detector at radius
  $300\km$. The range of $\Ye$ is divided into bins of width $0.01$. The
  histograms are normalized over the total ejected mass $M_{\rm ej}$. The
  left panel refers to the LS220 EOS simulations, the middle one to the
  DD2 EOS and the right one to the SFHO EOS; different line types mark
  binaries with different masses and mass ratios.}
\label{fig:ye-distro}
\end{center}
\end{figure*}

Figure~\ref{fig:mass-eject} shows that the amount of ejected material is
in the range $0.5 - 1 \times 10^{-3} \Msol$, with two exceptions. The
first is the binary \texttt{SFHO-M1.45}, which collapses immediately to a
black hole and results in very little material ejected (almost an order
of magnitude less), as most is accreted onto the black hole. Conversely,
the binary \texttt{SFHO-M1.35} model ejects a significant amount of
material when compared with the other models. Also this binary collapses
to a black hole around $9\,\ms$ (see star symbol in
Fig.~\ref{fig:mass-eject}) and since the SHFO EOS is a rather soft one,
this HMNS is the most compact we have simulated. Under these conditions,
it is natural that the larger compressions attained will lead to stronger
shock heating and hence to a larger dynamical mass ejection.

A measurement of the ejected mass that is alternative to that contained
in Eq.~\eqref{eq:m_ej} consists in evaluating a volume integral of the
rest-mass density of the unbound material over the entire computational
domain \cite{Hotokezaka2013, Sekiguchi2015, Lehner2016, Dietrich2017},
\ie
\begin{equation}
M_{\rm ej}(t) = \int \rho_{*} W \sqrt{\gamma} \dd^{3}x.
\label{eq:vol}
\end{equation}
As a cross-check we have employed this measurement for model
\texttt{LS220-M1.35} and found that $M_{\rm ej}(t)$ in this case is
obviously not a monotonically increasing function of time, but reaches a
maximum of $M_{\rm ej} = 0.80\times 10^{-3} \Msol$. This measurement
differs only of $4\,\%$ with that obtained via Eq.~\eqref{eq:m_ej},
demonstrating the robustness of our mass ejection and that the $300\,\km$
measurement radius is the most robust choice. In addition, the downside
of the use of Eq. \eqref{eq:vol} is that because of the finite size of
the domain, material that reaches the outer boundary is no longer include
in the calculation and causes the total ejected mass to decrease. Due to
this, we have evaluated Eq.~\eqref{eq:vol} at $\sim3\,\ms$ after merger
where it reaches a maximum and thus introducing some level of
arbitrariness in the evaluation of the integral. This specific
arbitrariness does not arise with the flux-integral method
\eqref{eq:m_ej}, which is integrated over all time, but where a choice
needs to be made for the extraction radius.

Finally, we note that our measured values of the ejected masses are
systematically smaller than those reported in Ref.~\cite{Sekiguchi2015}
for the same masses and EOS. This is likely due to the neutrino treatment
employed here and to the fact that more-sophisticated M1-scheme with
heating, such as that used in Refs.~\cite{Sekiguchi2015, Sekiguchi2016},
can allow for material to be more energetic and hence to become more
easily unbound \cite{Foucart2015}. On the other hand, our measurements
agree with those of Ref.~\cite{Lehner2016}, where a similar leakage
approach was employed; at the same time, the preliminary use of an
M0-scheme as that used in Ref. \cite{Radice2016} is insufficient to
explain this difference in the ejected mass. Finally, since the amount of
the ejected material depends also on the specific properties of the
computational infrastructure [\eg the location of the extraction
radius\footnote{In our calculations we have found that the difference
between the sphere at $300\km$ and a sphere at $740\km$ is about $30\%$
irrespective of EOS. Although the sphere further away has a slightly
higher ejected mass, the properties of the fluid are very close to
atmosphere at these radii and should be avoided.} in Eq. \eqref{eq:m_ej},
or the size of the computational domain in Eq. \eqref{eq:vol}] only a
direct comparison of the various neutrino-transport schemes within the
same code can quantify the variance of the ejected matter on the neutrino
treatment or the numerical specifications.

\subsection{Electron-fraction distributions}
\label{sec:ye}

The electron fraction is an important ingredient to determine the
r-process nucleosynthesis yields since $\Ye$ is effectively a measure of
how many free neutrons are available. Typically, low-$\Ye$ environments,
\ie with more free neutrons, favour a robust r-process and yield a higher
fraction of heavier elements while in high-$\Ye$ regimes, \ie with less
free neutrons, the production of very heavy elements tends to be
suppressed. Differences in $\Ye$ also correspond to potential differences
in the properties of the resulting kilonova signal, due to the efficient
production (or lack thereof) of high-opacity elements such as
lanthanides. In particular, the so-called ``blue'' kilonovae (\ie peaking
at higher frequencies, in the optical band) are possible in environments
with $\Ye\gtrsim0.25$ and ``red'' kilonovae (peaking in the infrared) in
environments with $\Ye\lesssim0.25$ \cite{Metzger2016,Metzger2017} (we
will discuss the angular distributions of the thermodynamical quantities
and their impact on the kilonova in Secs. \ref{sec:kilonova} and
\ref{sec:morphology}).

\begin{figure*}
\begin{center}
\includegraphics[width=2.0\columnwidth]{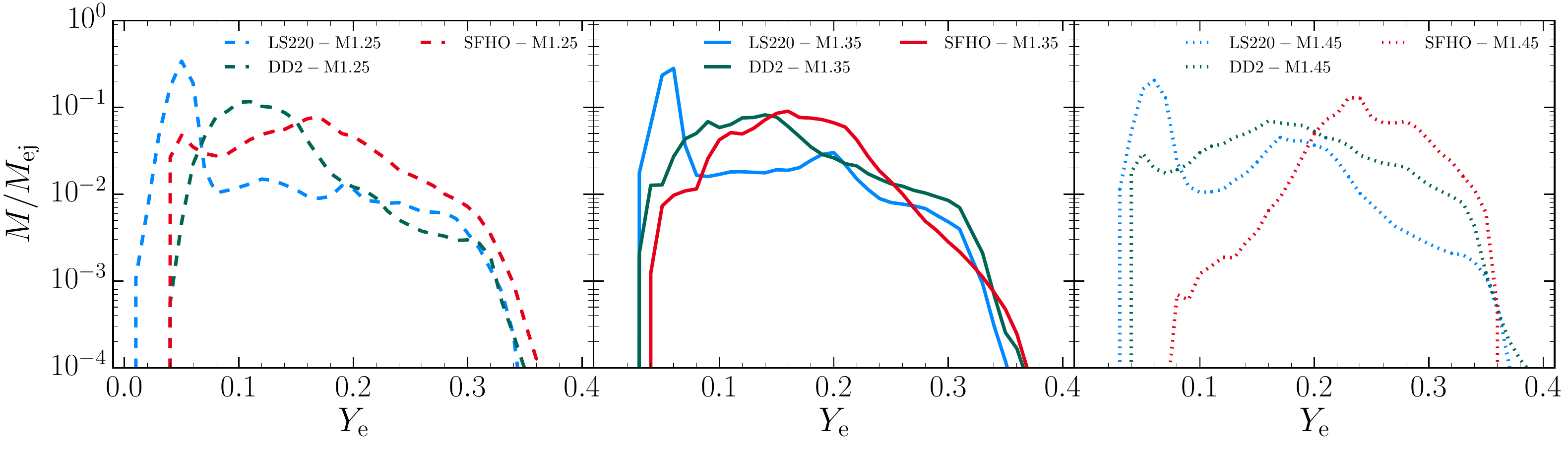}
\caption{Distributions of the ejected mass fraction as function of the
  electron fraction $\Ye$, as measured by a detector at radius
  $300\km$. This is the same as Fig.~\ref{fig:ye-distro}, except that the
  curves are here grouped by mass configuration rather than EOS so as to
  highlight the dependence on the latter. For clarity, unequal-mass
  binaries are not shown.}
\label{fig:ye-distro-mass}
\end{center}
\end{figure*}

Figure \ref{fig:ye-distro} shows histograms of the mass distribution of
the ejected matter over the electron fraction for all 12 simulations, as
computed from the data relative to our fiducial detector at radius $300
\km$; different panels refer to different EOSs, while the various lines
refer to the different binaries we have evolved. In practice, each patch
into which the detector sphere is subdivided, the local electron fraction
value is recorded and the local amount of ejected mass is estimated.
These values are then integrated over time up to $T_f$ to produce the
mass ejected along with the corresponding $\Ye$; the resulting $\Ye$
range is divided into bins of width $0.01$ and the unbound mass of each
patch at each time is assigned to a bin according to its corresponding
value of $\Ye$, thereby generating the histograms shown in
Fig.~\ref{fig:ye-distro}.

Irrespective of the EOS and mass configuration of the runs, common
qualitative features emerge. For all EOSs, the ejected mass is
distributed in a range of $\Ye$ varying from approximately $0.04$ up to
$0.4$, peaking at $\Ye \lesssim 0.2$. The only exception is the
\texttt{SFHO-M1.45} model, which ejects little material due to black hole
formation and whose distribution peaks at higher values of $\Ye$. This
spread of the electron fraction over a wide range is due to the inclusion
of a neutrino treatment, which causes the number of electrons to change
due to weak interactions. Failure to take such interaction into account
would result in a very different distribution, sharply peaked at very low
values of $\Ye$, \ie pure neutron matter (see, \eg
Ref.~\cite{Radice2016}).

More in detail, the LS220 runs (left panel) exhibit very similar
distributions for all mass configurations, peaking at approximately
$\Ye=0.05$ with a secondary peak at $\Ye\approx0.2$ before sharply
dropping off at electron fraction values of $\Ye\gtrsim0.3$. The
distributions of the DD2 (middle panel) also all exhibit a similar
behaviour, with a sharp increasing at $\Ye\sim 0.05$ before broadening out
with a sharp drop around $\Ye\sim0.3$. Finally, the distributions of the
SFHO runs (right panel) exhibit a somewhat different behaviour, although
spanning a similarly broad range in $\Ye$. The main differences in this
case are the tail of the distribution at higher values of the electron
fraction. In all cases, most of the ejected matter is found at low values
of the electron fraction, \ie it is very neutron-rich, which suggests a
robust r-process in all of the cases considered.

This conclusion is also supported by Table~\ref{table:results}, where the
average values $\langle\Ye\rangle$ of the electron fraction are reported
for all 12 runs. The averages are computed over the mass/electron
fraction histograms of Fig.~\ref{fig:ye-distro}. As can be seen in all
simulations, the average value of the electron fraction in the ejecta is
approximately $0.15$ or lower, indicating on average a very neutron-rich
environment. The only exception is model \texttt{SFHO-M1.45} where
$\langle\Ye\rangle=0.24$.

\begin{figure*}
\begin{center}
\includegraphics[width=2.0\columnwidth]{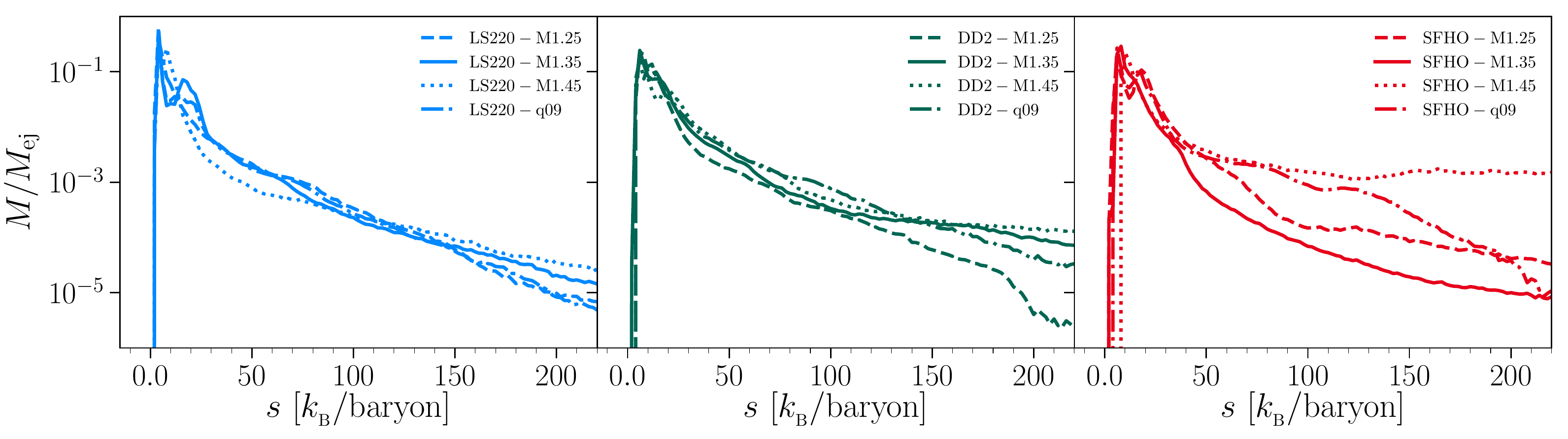}
\caption{The same as in Fig. \ref{fig:ye-distro} but for the specific
  entropy $s$. The range of $s$ is divided into bins of width $2\,\kB$
  and the histograms are normalized over the total ejected mass.}
\label{fig:s-distro}
\end{center}
\end{figure*}

In Fig.~\ref{fig:ye-distro-mass}, to help the comparison of the results
across different EOSs the distributions are arranged according to the
total mass of the BNS (the unequal-mass cases are excluded) instead of
EOS in Fig.~\ref{fig:ye-distro}. In all panels, there is a noticeable
trend in the distributions of $\Ye$, which is most evident in the
$1.25\Msol$-case (left panel), where $\langle \Ye\rangle = 0.08$, $0.13$,
and $0.14$ for LS220, DD2, and SFHO EOSs, respectively. This increase in
$\Ye$ is expected when considering that neutrino interactions depend
strongly on the temperature. The average entropy (see
Sec.~\ref{seq:entropy}) of these simulations is $10.3$, $13.9$, and
$15.6\, \kB$ respectively. Entropy is related to temperature and the
higher the entropy, the higher the average temperature, \cf
Fig.~\ref{fig:ye-distro}, and hence more free neutrons are converted into
neutrinos through positron capture, increasing $\Ye$. 

This effect is also related to the compactness of the object, albeit this
relation should be treated carefully. SFHO is the softest EOS, which
leads to the most compact objects. This results in higher temperatures
during the merger, which causes an increase in the neutrino reactions,
which decreases the number of neutrons and as expected has the highest
average $\Ye$. In contrast, from Table \ref{tab:models} one would expect
that because LS220 is more compact than DD2, LS220 should have a higher
average $\Ye$ and entropy when the opposite is this the case. This
difference is due to compactness being a property calculated for cold
beta-equilibrium where the effects of composition are minimal.  As
discussed in Sec.~\ref{sec:ID}, the LS220 does not include light nuclei
which can change the composition and the neutrino interactions so this
seemingly non-monotonic relation between compactness and average $\Ye$
arises from different constructions of the EOS. When comparing DD2 and
SFHO and excluding LS220, there is a clear monotonic relationship between
$\mathcal{C}$ and $\Ye$.

When comparing our results with that of simulations with similar initial
data, there is a disagreement with computed values of the electron
fraction. For example, for the \texttt{DD2-M1.35} model with our
measured value of $\langle \Ye\rangle = 0.14$, the authors of
Ref.~\cite{Sekiguchi2015} report $\langle \Ye\rangle = 0.29$ with an
M1-scheme independent of resolution and $\langle \Ye\rangle = 0.26$ with
a leakage scheme with a resolution of $200$ m. However, a similar
distribution in $\Ye$ is observed in Refs.~\cite{Palenzuela2015,
Lehner2016}, which use a similar leakage scheme to the one used here.

\subsection{Specific-entropy distributions}
\label{seq:entropy}

The next thermodynamic quantity we consider is the distribution of the
ejected material over the entropy per baryon $s$. The specific entropy is
important in r-process nucleosynthesis as it impacts the neutron-to-seed
ratio, with high initial neutron-to-seed ratios favouring the production
of heavy nuclei during the r-process nucleosynthesis even at relatively
high electron fractions. In particular, in radiative environments such
as those accompanying the ejected matter, the specific entropy will scale
with the temperature as $s \propto T^3$, so that the shock-heated (and
hotter) part of the dynamical ejecta will exhibit higher entropies. In
turn, because the seed nuclei will be photodissociated at high
temperatures, a higher specific entropy will increase the neutron-to-seed
ratio and thus r-process nucleosynthesis. In contrast, the cold, tidal
dynamic ejecta, and which dominates the unbound matter in Newtonian
simulations, (see, \eg Refs.~\cite{Freiburghaus:1999, Rosswog1999,
Rosswog2002}) usually exhibit low entropy, but extremely neutron-rich
material \cite{Korobkin2012}. The distributions of the specific entropy
computed with the same procedure as the electron fraction distribution in
the previous section, is shown in Fig.~\ref{fig:s-distro}, while the
average values $\langle s\rangle$ are reported in
Table~\ref{table:results}.

Again, we observe many EOS-independent qualitative features. First, for
all EOSs, the mass distribution peaks at $s \approx 2\,\kB$, while a fast
decay is visible towards higher entropies. In the case of the binaries
with the DD2 EOS (middle panel), the qualitative behaviour of different
mass configurations is similar up to approximately $s\simeq100\,\kB$. At
larger entropies, the \texttt{DD2-M1.25} binary has a more rapid drop-off
and there is very little material that reaches higher entropies. In
comparison, the remaining models exhibit similar behaviour with a
flattening of the curve at higher entropies. The average entropy value is
in all four cases $\langle s\rangle\approx 15\,\kB$. Second, all of
binaries with the LS220 EOS (left panel), show a very similar qualitative
behaviour among themselves and strong analogies with the DD2 binaries. In
particular, the distributions show a rapid increase in entropy at around
$2\,\kB$ (for the $1.45\Msol$ binary this peak is at around $8\,\kB$ and
is 4 times smaller), with an additional second peak at $20\,\kB$ for the
$1.35\Msol$ case that is not present in the other masses. For all
masses, there is a rapid decrease in specific entropy, with average
entropies that are slightly lower than the DD2 and SFHO binaries and with
a smaller spread between the values, being approximately $s\sim 11
\,\kB$.

Finally, the simulations with binaries having the SFHO EOS (right panel)
show a similar qualitative behaviour with the other runs, at least at low
entropies. The distributions peak at about $ 5\,\kB$ and a rapid drop
follows, although different binaries show different fall-offs at around
$50\,\kB$. In the \texttt{SFHO-M1.25} case, the distribution begins to
decrease less rapidly at higher entropies while the \texttt{SFHO-M1.35}
model shows the fastest decrease. This is in contrast to the DD2 and
LS220 simulations (where the specific entropies correlate with the
initial masses of the stars) and is reflected in the average values of
the specific entropy, with the \texttt{SFHO-M1.25} model having $\langle
s\rangle=15.6\,\kB$, while \texttt{SFHO-M1.35} a smaller value of
$\langle s\rangle=12.7\,\kB$. Lastly, The average specific entropy of the
\texttt{SFHO-M1.35} binary is almost twice as large, likely due to the
fact that the small amount of ejected matter has been efficiently heated
on account of its rarefaction. While somewhat puzzling, this
non-monotonic behaviour of the specific entropy with the SFHO binaries is
likely due to the comparative softness of this EOS, which enhances the
nonlinearity associated with shock-heating effects.

Indeed, as with the electron-fraction distributions, the average entropy
tends to increases with the softness of the EOS\footnote{Taking into
consideration the caveats at the end of Section~\ref{sec:ye}.}, being the
highest for the softest EOS, \ie SFHO. For example, concentrating on the
$1.25\Msol$ binaries, $\langle s\rangle = 10.3,\,13.9,\,15.6\,\,\kB$, for
the LS220, DD2, SFHO EOSs, respectively. This dependence is not
particularly surprising as softer EOSs produce a higher temperature and
the temperature is directly related to the specific entropy. This
relation holds for almost all cases, even when including the low-mass
ejecta of \texttt{SFHO-M1.45}; the only exception is offered by the
\texttt{SFHO-M1.35} binary, where this discrepancy is likely due to there
being at least $5$ times as much ejecta as the other binaries.

\subsection{Ejection-velocity distributions}
\label{sec:velocity}

\begin{figure*}
    \begin{center}
    \includegraphics[width=2.0\columnwidth]{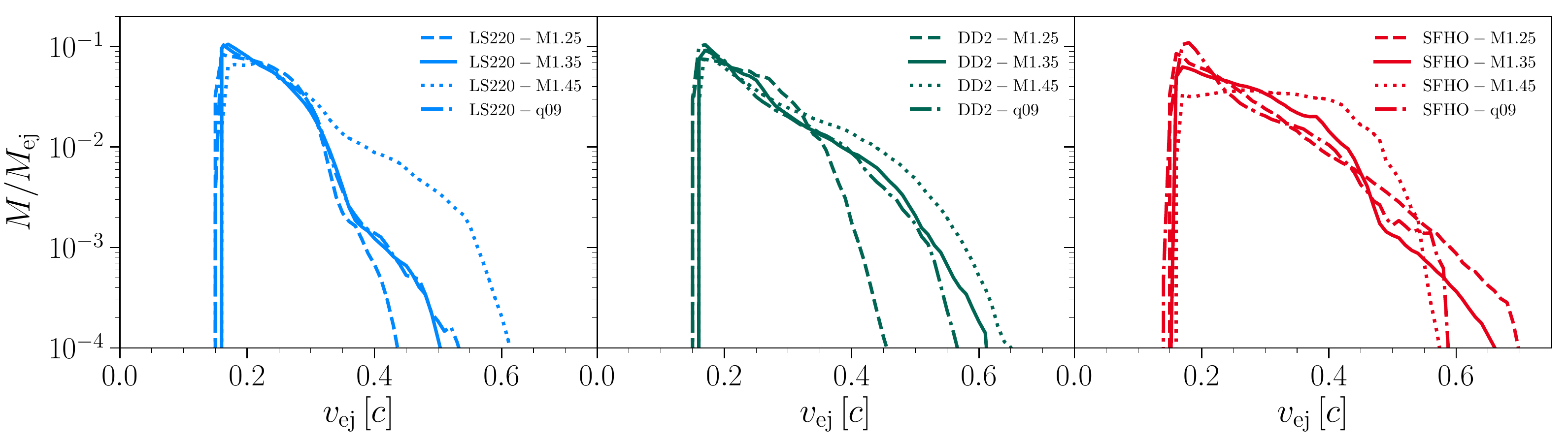}
    \caption{The same as in Fig. \ref{fig:ye-distro} but for the ejecta
      velocity $v_{\rm ej}$. The range of $v_{\rm ej}$ is divided into
      bins of width $0.05$ and the histograms are normalized over the
      total ejected mass.}
    \label{fig:v-distro}
    \end{center}
\end{figure*}

Figure \ref{fig:v-distro} reports the velocity distributions of the
ejecta computed in full analogy with the electron-fraction or
specific-entropy distributions presented in the previous two
sections. Note that unlike, \eg Ref. \cite{Radice2016}, we here
distinguish between the velocity of the ejected material $v_{\rm ej}$ as
measured in the simulation and that of the ejecta at spatial infinity
$v_{\rm {inf}}$. In particular, we compute $v_{\rm ej}$ directly from the
Lorentz factor $W$, \ie $v_{\rm ej} =\left[(W^2-1)/W^{2}\right]^{1/2}$,
where we assumed that the detectors are are sufficiently far away from
the merger product so that the Minkowski metric holds. As discussed in
Ref.~\cite{Bovard2016}, this is a rather good approximation since it was
shown there that the ejected matter moves essentially radially and there
is only a subdominant velocity component in the angular directions, hence
$v^{2} \approx v_{r}^{2}$, which enables us to compute $v_{\rm ej}\simeq
v^r$ from $W$. An obvious consequence of distinguishing between $v_{\rm
  ej}$ and $v_{\rm {inf}}$ is that our values of the ejecta velocities
are systematically higher than in Ref.~\cite{Radice2016}.
 
Again, Fig. \ref{fig:v-distro} reveals that every simulation exhibits
similar qualitative behaviour. The ejecta velocity is never lower than
$0.15\,c$; the bulk of the matter has velocities of $v_{\rm ej}\approx
0.25\,c$, and at higher velocities of $v_{\rm ej} \gtrsim 0.6\,c$ the
mass distribution quickly drops to zero. Table~ \ref{table:results}
reports the average velocity $\langle v_{\rm ej}\rangle$ for all the
runs. A trend clearly emerges from our data, with the higher-mass
configurations systematically producing higher-ejecta velocities. More
precisely, the ejecta velocity appears to be tightly correlated with the
compactness of the neutron stars involved in the merger (\cf
Table~\ref{tab:models}). Also in this case, this trend is not
particularly surprising since higher-mass configurations result in more
compact starts, which in turn experience stronger torques and more
efficient shock heating.

In Table~\ref{table:results}, the column denoted by $\langle
v_\infty\rangle$ shows estimates of the ejecta velocity at infinity,
which is achieved in the homologous expansion phase. This velocity is
used in our approximate model of kilonova emission (see Sec.
\ref{sec:kilonova}) and is computed assuming a ballistic radial motion
from $r=300\,\km$ to infinity in the spherically symmetric
gravitational field of an object with the same ADM mass of the BNS system
under consideration, \ie 
\begin{align}
\label{eq:v_inf}
    \sqrt{1-\frac{2M_{_{\rm ADM}}}{r}}
    \frac1{\sqrt{1-\langle v_{\rm ej}\rangle^2}}
    =
    \frac1{\sqrt{1-\langle v_\infty\rangle^2}}\,.
\end{align}
In the Newtonian limit, $M_{_{\rm ADM}} = M$ and expression
\eqref{eq:v_inf} simply reduces to the familiar energy conservation
equation: $\tfrac{1}{2}\langle v_{\rm ej}\rangle^2 - {GM}/{R} =
\frac12\langle v_\infty\rangle^2$.

\section{r-process nucleosynthesis}
\label{sec:abundances}

This section is dedicated to a discussion of our analysis of the
nucleosynthesis of r-process material taking place for the matter that
has been ejected dynamically in the merger and is unbound. In particular,
we will concentrate on the optimal selection of the tracers, on how
nucleosynthesis varies with the specific entropy of the ejected matter
and on those behaviours that are essentially independent of the EOS.

\subsection{Tracer-input comparison}

In Sec. \ref{sec:traceroutflow} we introduced a method to associate a
mass to the otherwise massless tracers. Here, we introduce two additional
tracer-selection criteria (together with the unboundness criterion
already discussed and which is always enforced) and the corresponding
procedures to associate a mass to the tracers; we then compare the impact
that this different selection strategies have on the final
nucleosynthesis yields.

We recall that the first criterion, introduced in
Sec. \ref{sec:traceroutflow}, consists in considering from a given
simulation all tracers that are unbound, associate to each of them a mass
by locally integrating a mass flux through a sphere of given radius as in
Ref.~\cite{Bovard2016} (in our case, the sphere is the fiducial outflow
detector at radius $200\,\Msun$), then sum the nucleosynthesis yields
from all tracers using the corresponding mass as weight to recover the
final abundance pattern. Since the total number of unbound tracers in one
of our simulations can reach several thousands ($40,\!000$ being a
typical value), this approach involves the post-processing and
book-keeping of many tracer trajectories, thus becoming computationally
rather costly. For this reason, we develop the alternative selection
criteria described below.

The second tracer-selection criterion consists instead in considering the
distributions of the ejected mass as a function of the electron fraction
presented in Sec. \ref{sec:ye} and in drawing one representative, unbound
tracer from each bin. Given the bin width of $\Delta Y_{\rm e}=0.01$,
this results in about $40$ tracers for every simulation, a reduction of a
factor of a thousand with respect to the first criterion. In this
approach a mass is then associated to each tracer by assigning to it the
mass of the bin it was drawn from. We refer to this procedure as to the
``1D'' criterion, since the tracers are drawn from a 1D distribution.

\begin{figure}
    \begin{center}
    \includegraphics[width=\columnwidth]{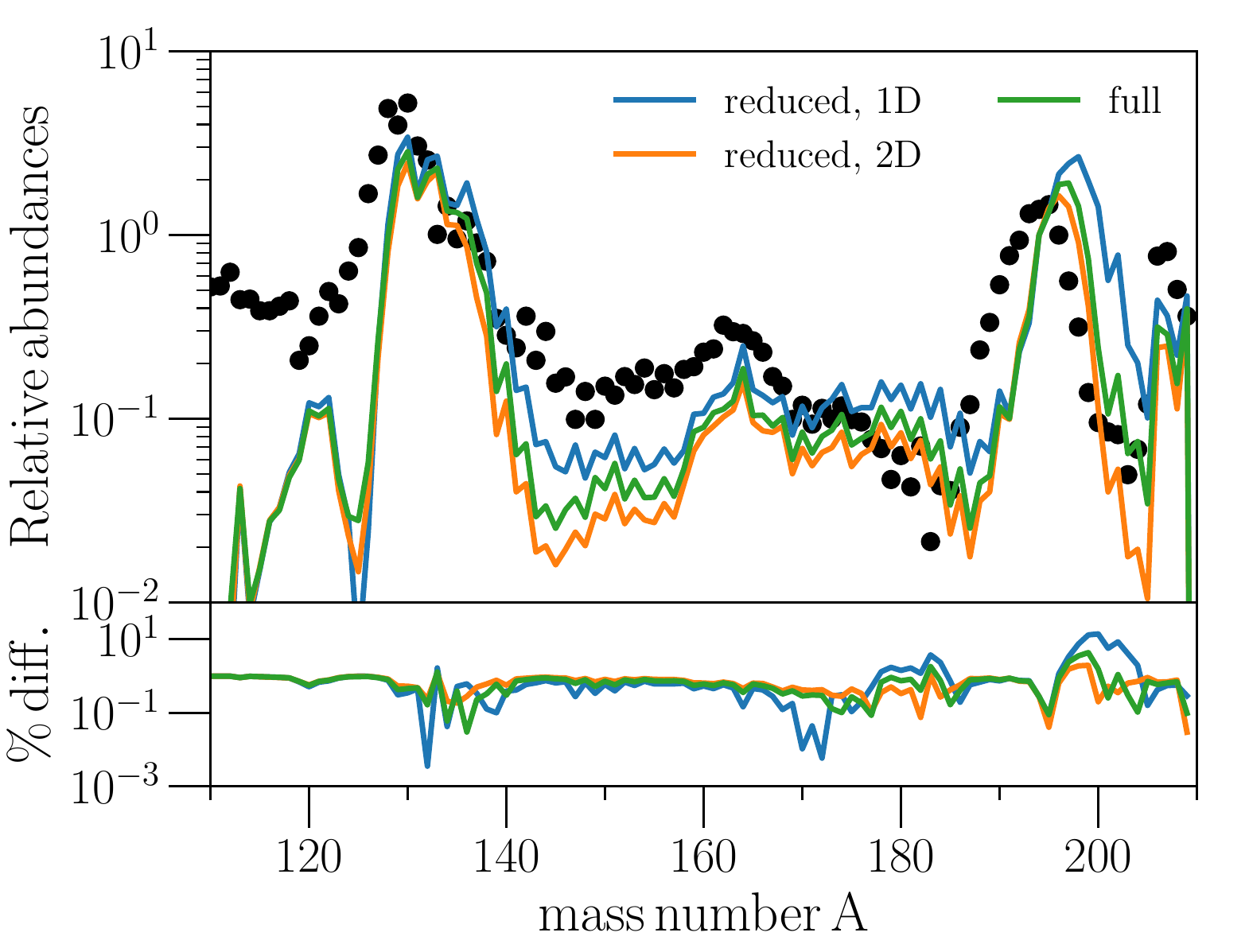}
    \caption{Comparison of the relative abundances $Y_i$ of the r-process
      as function of the mass number $A$ for the three tracer selection
      criteria. In blue, the abundances produced by the ``1D'' criterion;
      in orange the ones produced by the ``2D'' criterion; in green the
      abundances obtained considering all unbound tracers.  The black
      filled circles indicate the solar abundances. In the bottom panel,
      the relative differences of the three lines from the solar
      abundances are shown.}
    \label{fig:tracer-comp}
    \end{center}
\end{figure}

We finally consider a third selection criterion, essentially an improved
version of the 1D criterion. It consists in considering the ejected mass
histogram over both the electron fraction and the specific entropy; we
then draw one representative, unbound tracer from each bin, and associate
to it the mass of the bin it was drawn from. We refer to this procedure
as to the ``2D'' criterion, since the tracers are drawn from a 2D
distribution. For each simulation, this results in a total of roughly
$1,\!000$ tracers to be considered.

We show in Fig.~\ref{fig:tracer-comp} the results from the
nucleosynthesis calculations for the three selection criteria. We
restrict the comparison to one fiducial case, the binary
\texttt{LS220-M1.35}, and compare our final abundance pattern with the
solar one (filled circles), showing the relative difference to it in the
bottom panel. As can be seen, the original approach of considering all
unbound tracers reproduces quite well the solar abundances over the whole
range of mass numbers considered, as does the 2D criterion. The 1D
criterion instead shows significant deviations, especially around the
third peak (\ie $A \simeq 195$) and around the rare-earth peak (\ie $A
\simeq 165$). A posteriori, this is due to the fact that the 1D criterion
is systematically biased towards low-entropy tracers, which has a
significant impact over the final abundances, as we discuss in the next
section.

\begin{figure}
    \begin{center}
    \includegraphics[width=0.95\columnwidth]{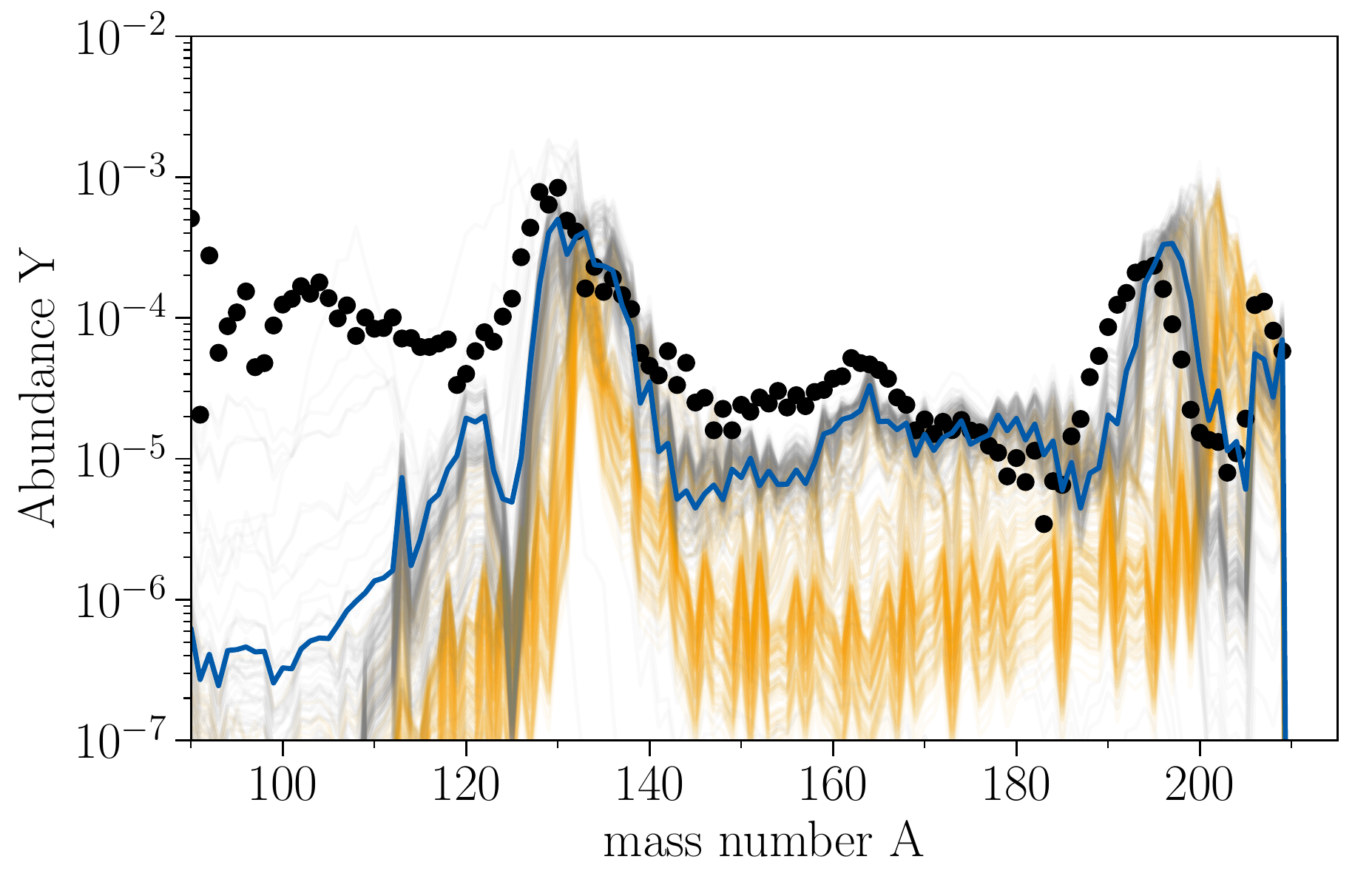}
    \caption{Final r-process abundances for all unbound tracers of the
      \texttt{LS220-M.1.35} binary. Gray lines are the yields for
      individual tracers with low entropies \mbox{$s<70\,\kB$}, and
      orange lines mark single tracers with high entropies
      \mbox{$s\geq70\,\kB$}. The mass-integrated nucleosynthesis yields
      are shown with a blue line, and the black filled circles show the
      solar abundances.}
    \label{fig:nucleosynthesis-ls220-m1.35-full}
    \end{center}
\end{figure}
\begin{figure*}
    \begin{center}
    \includegraphics[width=0.99\textwidth]{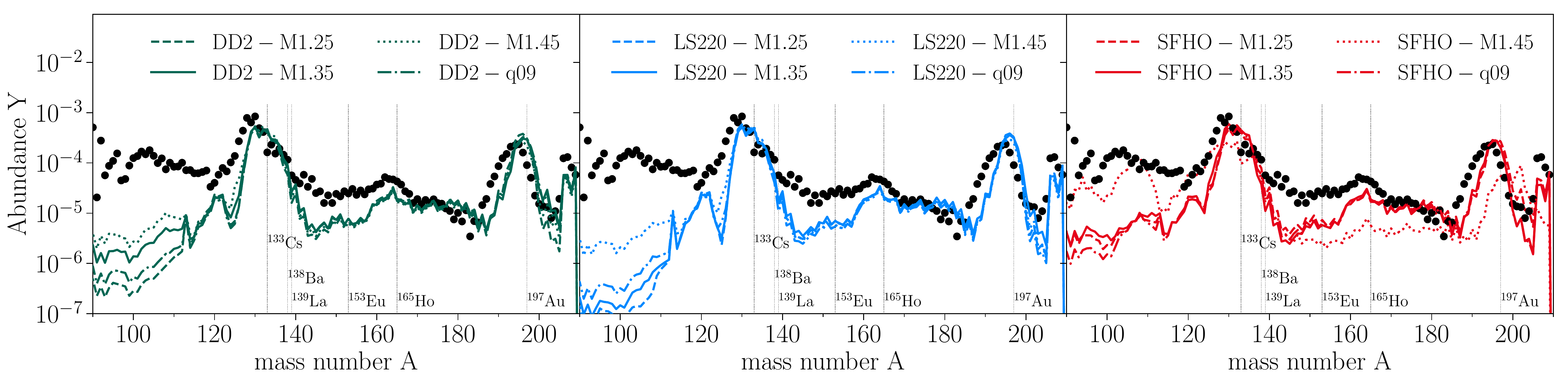}
    \caption{Final relative heavy-elements abundances for all the 12 BNS
      models as a function of mass number $A$. The abundances are
      normalized so that the total mass fraction is unity, while the
      different panels and lines refer to the various EOSs, masses and
      mass ratios, respectively (see legends). The black filled circles
      report instead the observed solar abundances, while the vertical
      lines mark a few representative r-process elements: ${}^{133}\rm
      Cs$, ${}^{138}\rm Ba$, ${}^{139}\rm La$, ${}^{153}\rm Eu$,
      ${}^{165}\rm Ho$, ${}^{197}\rm Au$.}
    \label{fig:nucleosynthesis-normalized}
    \end{center}
\end{figure*}

The 2D criterion is computationally much less expensive than considering
all unbound tracers, it allows for a simple and unambiguous definition of
the tracer mass, and yet it leads to an almost unbiased abundance
calculation. In the following, we adopt it as our preferred
tracer-selection criterion and compute all results with it, unless
otherwise stated.

\subsection{Heavy-element nucleosynthesis}
\label{sec:mass_nucleo}

Figure \ref{fig:nucleosynthesis-ls220-m1.35-full} illustrates the
nucleosynthesis results for all $\sim 40,\!000$ unbound tracers of the
representative simulation of the \texttt{LS220-M.1.35} binary. In
particular, we plot individual tracers with $s<70\,\kB$ in gray or
$s\geq70\,\kB$ in orange, respectively, alongside the mass-integrated
abundances (blue line). As a consequence of the relatively low electron
fractions for most of the ejecta (\ie with $\Ye\approx0.1$; see Fig.
\ref{fig:ye-distro} and Table~\ref{table:results}), we obtain that for
each tracer the strong r-process component from the second to the third
r-process peak are well reproduced. At the same time, we find that the
entropy distribution of the ejecta gives rise to specific features in the
abundances pattern. More specifically, the low-entropy component (\ie
$s<70\,\kB$) leads to the pattern that is observed in the neutron-rich
ejecta of Newtonian simulations. On the contrary, the high-entropy (\ie
$s\geq70\,\kB$) part of the ejecta, which carries only about $6\,\%$ of
the total ejected mass, has a nucleosynthesis pattern with a shifted
second and third peak. Additionally, it shows diminished abundances in
the rare-earth region, and effectively fills the gap between third
r-process peak and elements in the Lead region. We note that the
abundance pattern of these tracers is very similar to the ``fast'' ejecta
found by the authors of Ref.~\cite{MendozaTemis2015}. While we do not
find them to expand faster in the beginning, their unusual abundance
distribution can be traced back to an extremely high initial
neutron-to-seed ratio $Y_{\rm n}/Y_{\rm seed} \gtrsim 1,\!000$ and
comparably low initial densities $\rho \lesssim 10^9\gcm$. Due to the
enormous amount of neutrons at low densities, the seed nuclei require
substantially more time to incorporate the neutrons, delaying the
freeze-out time (\ie the time when $Y_{\rm n} / Y_{\rm seed} = 1$). In
fact, the time window for the r-process to occur in this minority of
ejected material is $\gtrsim 100\s$ instead of $\lesssim 1\s$. Moreover,
the r-process runs along a path much closer to the valley of stability
for these tracers, such that the magic neutron numbers are reached at
higher mass numbers, and the abundances settle down for a pattern in
between s-process and r-process.

Figure \ref{fig:nucleosynthesis-normalized}, on the other hand, reports
the final heavy-elements relative abundances for all of the 12 BNS models
outlined in Table~\ref{tab:models} and shows them as a function of the
mass number $A$. As in previous figures, the different panels refer to
the different EOSs considered and the various binaries are represented
with lines of different types. Furthermore, The results are normalized to
have a total mass fraction of 1 and shown with filled circles are the
scaled solar system r-process abundances. Clearly, in all cases, a
successful r-process is obtained, leading to the production of the whole
r-process pattern from the second (\ie $A \sim 130$) to the third (\ie $A
\sim 195$) peak.

\begin{figure*}
    \centering
    \includegraphics[width=2.0\columnwidth]{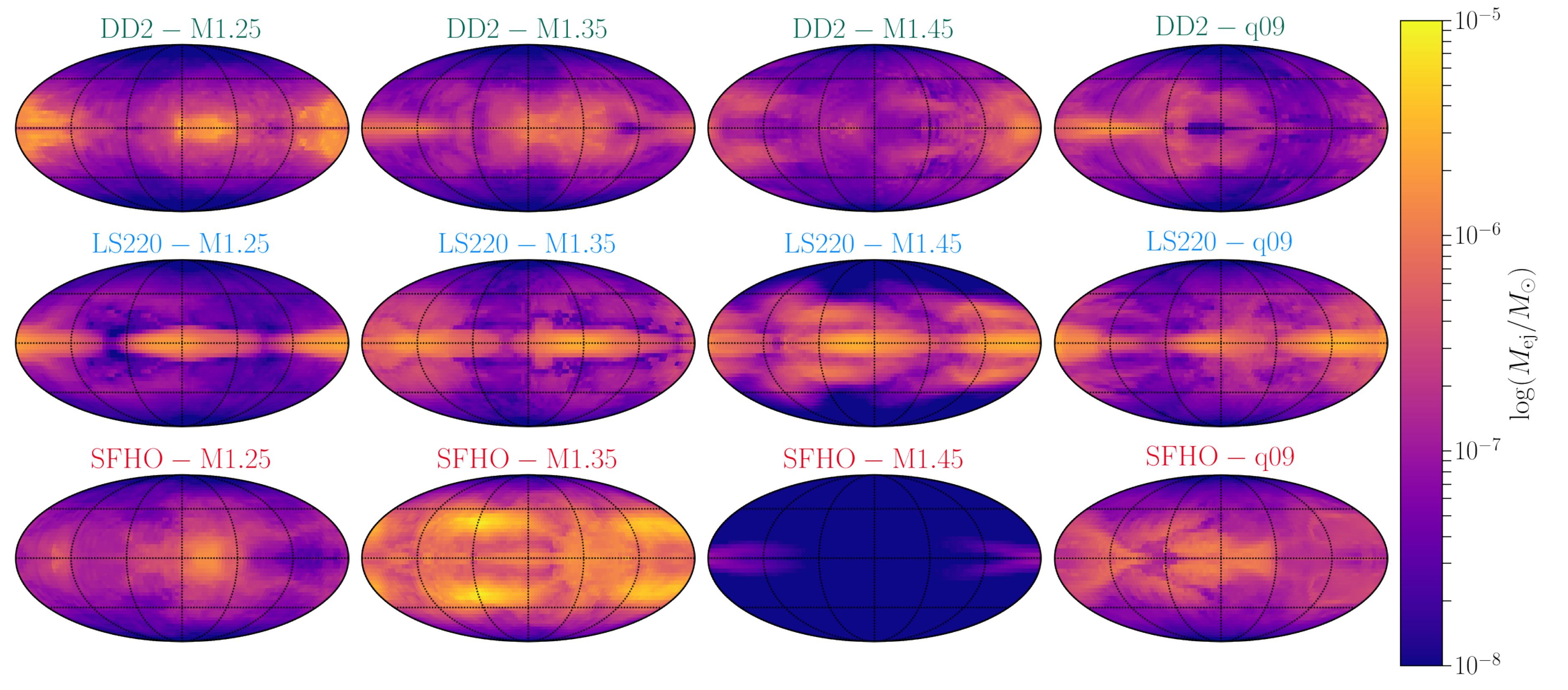}
    \caption{Angular distribution of the ejected mass at the final time
      for the various binaries, with the different rows referring to the
      different EOSs considered.}
    \label{fig:mass2D}
\end{figure*}

However, there are different admixtures due to the different
electron-fraction distributions of the ejected material as detailed in
Sec. \ref{sec:ye}. For the equal-mass binaries, in particular, we observe
a tendency of slightly enhanced abundances below the second r-process
peak with increasing mass of the neutron stars. This is because more
massive BNS systems have a higher electron fraction on average.
Furthermore, the contributions from tracers with high initial
neutron-to-seed ratios enhance both the second r-process peak and the
region with $A \approx 200$ in all cases. The most extreme example is the
\texttt{SFHO-M1.45} binary, which immediately collapses to a black hole
after merger, ejects very little mass and with a comparatively high
electron fraction. As a result, the part of ejected material with low
specific entropy leads to nuclei that mainly have mass numbers with $A
\lesssim 130$, while the material with high specific entropy -- and thus
high neutron-to-seed ratios - dominates the final abundances beyond the
second r-process peak, leading to an enhanced abundance for $A \gtrsim
200$. The distinctive features observed in the final abundances in the
case of the \texttt{SFHO-M1.45} binary opens therefore the prospect of
using the chemical yields either as a confirmation of the prompt
production of a black hole after the merger, or as an indication of this
process in the case in which the post-merger gravitational-wave signal is
not available.

All things considered, the most striking result shown in
Fig. \ref{fig:nucleosynthesis-normalized} is the excellent and
\emph{robust} agreement of the various abundance patterns,
notwithstanding the fact that they have been obtained using different
combinations of EOSs and neutron-star masses. While this agreement might
be partly aided by our simplified neutrino treatment, this result not
only confirms the robustness of the r-process yields from BNS mergers
already noted in the literature, but it also shows how the uncertainties
associated in modelling the microphysics of BNS mergers have a very
limited impact on the nucleosynthesis. In fact, the spread in our
r-process patterns is much less than the one associated to uncertainties
in the nuclear-physics modelling of nuclei involved in the r-process, \eg
the choice of the fission fragment distribution \cite{Eichler2014} or the
nuclear mass model (see, \eg Refs.~\citep{Mumpower2015a,Martin2016}).

\section{Ejecta morphology and kilonova light curves}
\label{sec:kilonova}

\subsection{Angular distributions of ejected matter}
\label{sec:morphology}

The use of outflow detectors allows us to study, in addition to the
properties of the ejected material, the angular distribution of the
ejected material on the detector surface and hence virtually at spatial
infinity. Besides having an interest in their own right, anisotropies in
the distribution of the ejected matter could have important consequences
on the kilonova signal of a given binary configuration, and impact its
detectability. To the best of our knowledge, this is the first time that
an analysis of this type has been carried out.

\begin{figure*}
    \centering
    \includegraphics[width=2.0\columnwidth]{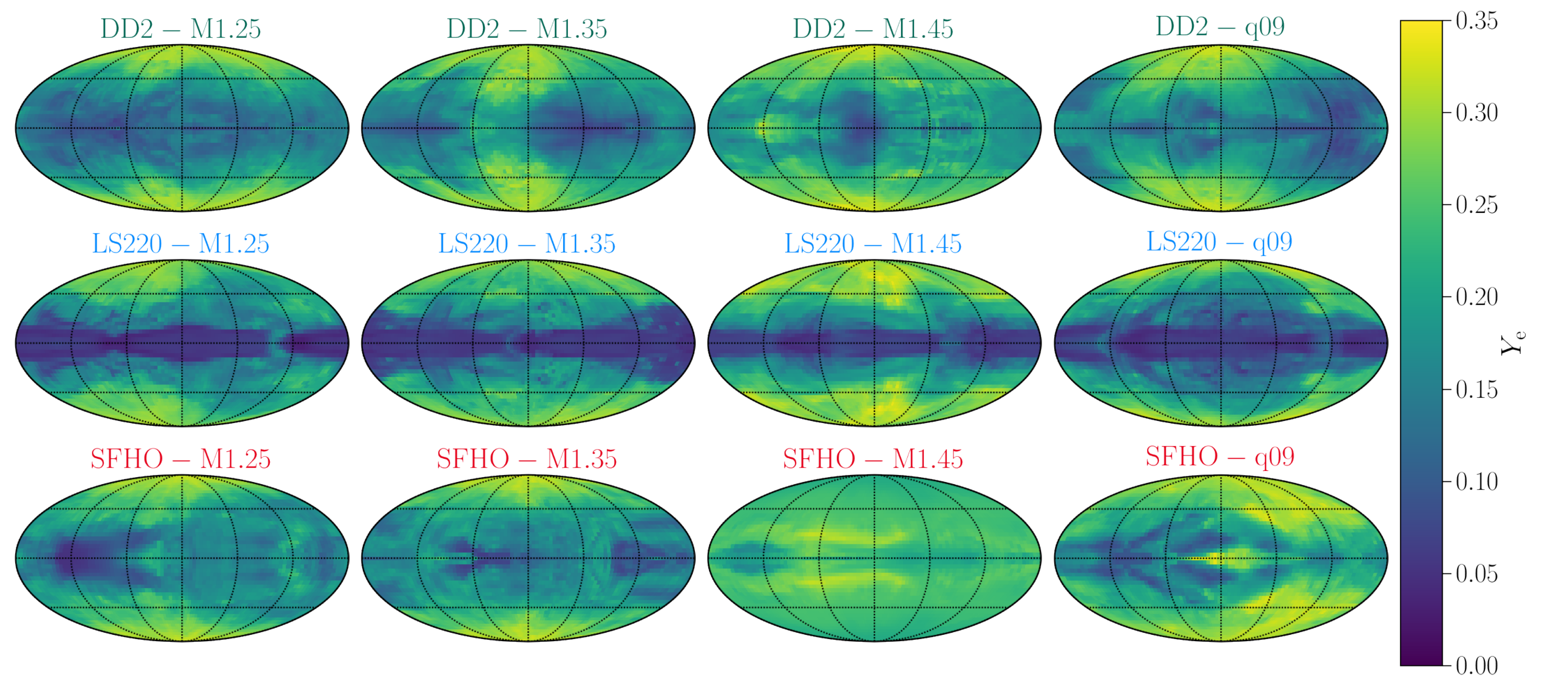}
    \caption{The same as in Fig. \ref{fig:mass2D} but for the electron
      fraction.}
    \label{fig:Y_e_bar2D}
\end{figure*}

In practice, we consider the angular distribution of ejected mass as
defined by \eqref{eq:m_ej}, where in this case however the integration
over the angular directions does not span the whole 2-sphere, but only a
single patch of the outflow detector. We also study the mass-averaged
distribution of the electron fraction, the specific entropy and the
ejecta velocity. Similarly to Eq.~\eqref{eq:m_ej}, these are defined as
\begin{align}
  \langle \chi \rangle
  := &{\int_0^{T_f} \int_{\Delta\Omega} \chi\, \rho_\ast W (\alpha
    v^r - \beta^r) S\, d \Omega\, d t}\ /& \nonumber \\
     & {\int_0^{T_f} \int_{\Delta\Omega}
    \rho_\ast W (\alpha v^r - \beta^r) S\, d \Omega\, d t}\,,
    \label{eq:morph}
\end{align}
where $\chi$ is any one of $\Ye$, $s$ or $v_{\rm ej}$, and the same
consideration as above applies to the integration over the angles.  In
this section we only focus on the results obtained for the angular
distribution of the ejected mass and the electron fraction. The
corresponding analysis in the case of the specific entropy and of the
ejecta velocity is reported in Appendix \ref{sec:add_morphology}.

In Fig.~\ref{fig:mass2D}, we report Mollweide projections of the outflow
detector relative to the time-integrated rest mass for all models.
Several observations are in order: First, the binary \texttt{SFHO-M1.45},
which immediately collapses to a black hole after the merger, is
immediately identifiable as there is close to no ejected matter in this
case. Second, it is clear that, in each binary, most of the mass is
ejected on the orbital plane, which is consistent with expectations that
the material ejected here is mostly of dynamical origin and is powered by
the torques in the system at merger (other types of ejecta, such as
neutrino/magnetically driven winds or ejecta from viscous heating could
display a more isotropic structure). Third, while concentrated at low
latitudes, the ejected mass is not uniformly distributed but shows
considerable anisotropies; this is simply due to the disruption flows
produced by the tidal torques and this concentrates the emission of
matter into rather small regions on the detector surface.  The only
binary that appears to evade this trend is \texttt{SFHO-M1.35}, which has
ejected also at latitudes as high as $\sim\,45^\circ$ and seems to peak
around $\sim\,30^\circ$.

Similarly, the distribution of the electron fraction $Y_{\rm e}$ is shown
in Fig.~\ref{fig:Y_e_bar2D}. It can immediately be appreciated how the
electron fraction tends to anticorrelate with the amount of ejected mass:
regions in which the ejected mass fraction is higher (such as the orbital
plane) tend to have very low $\Ye$ and vice-versa. This consistent with
the results of Sec.~\ref{sec:ye}, where most of the ejected mass was
shown to be very neutron-rich. On the other hand it can be seen that in
other regions, such as the poles, the material is very neutron-poor, but
has correspondingly low values of ejected mass. The evidence provided in
Fig.~\ref{fig:Y_e_bar2D} that matter ejected around the poles is less
neutron-rich (\ie with $\Ye \gtrsim 0.25$) suggests the possibility that
material there might undergo a less robust r-process, leading to a
suppressed production of lanthanides and thus to a lower opacity. This
bimodal anisotropy in the distribution of the electron fraction could
then lead to either a ``blue'' kilonova, \ie to a kilonova signal with a
comparatively strong optical component, if the line of sight is mostly
along the polar regions, or to a ``red'' kilonova, \ie to a kilonova
signal peaking in the infrared, if the line of sight is mostly along the
equatorial regions. 

\begin{figure}[!t]
    \centering
    \includegraphics[width=1.0\columnwidth]{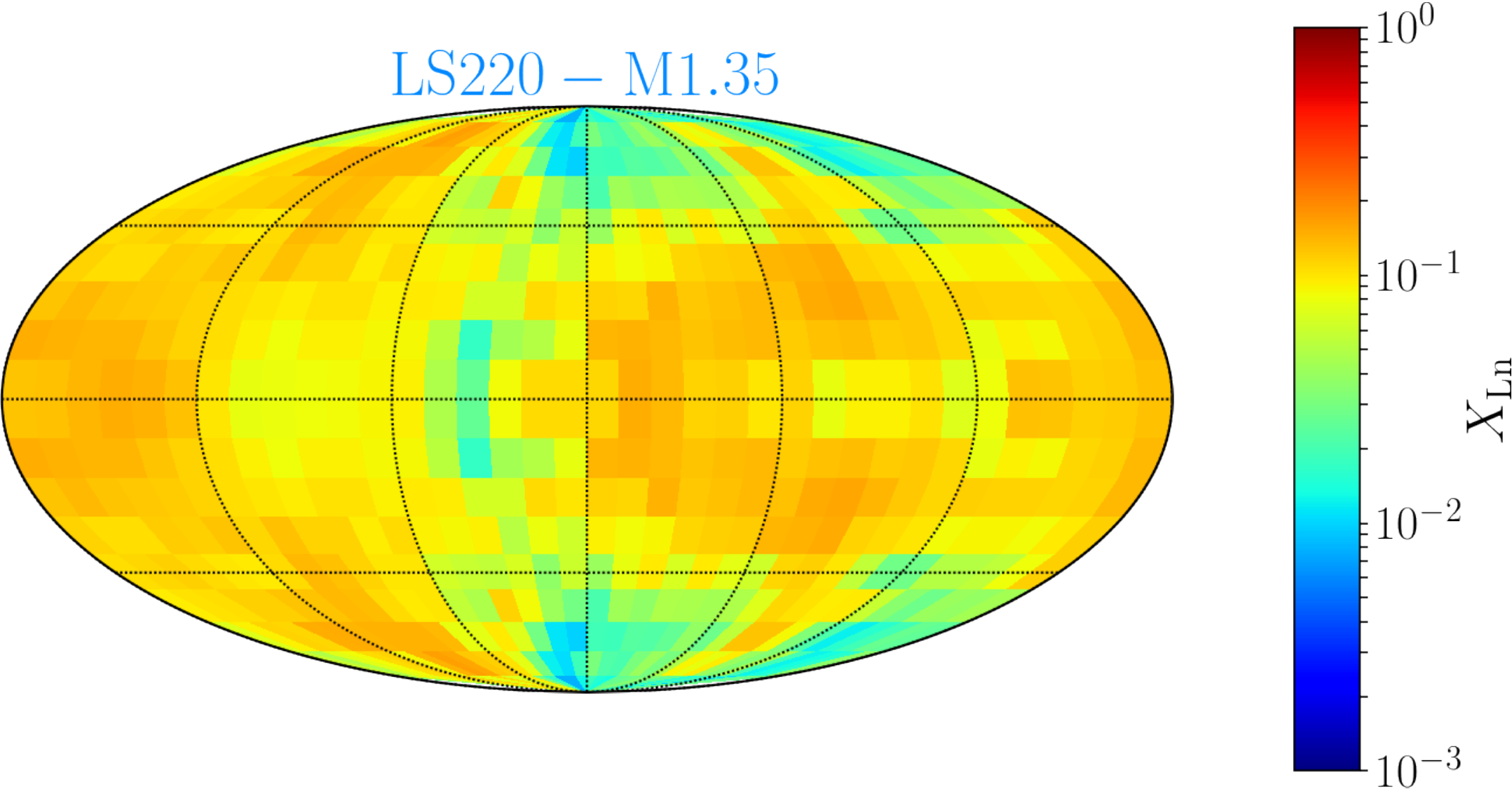}
    \caption{Angular distribution of the mass fraction of lanthanides in
      the representative case of the binary \texttt{LS220-M1.35}; the
      data refers to the final simulation time.}
    \label{fig:lanthanides2D}
\end{figure}

We have checked the plausibility of such a scenario by explicitly
computing the angular distribution of the lanthanides mass fraction in
the representative \texttt{LS220-M1.35} model. This has been obtained by
computing the lanthanides mass fraction of every unbound tracer in the
simulation and by plotting their location on the 2-sphere, as shown in
Fig.~\ref{fig:lanthanides2D}, where the lanthanides mass fraction values
have been averaged over patches of angular size $10^\circ \times
10^\circ$. As can be seen from the figure, even near the poles, the
lanthanides mass fraction is rather high, \ie $X_{\rm Ln} \approx
10^{-2}$. This is far larger than the generally accepted limit on this
value that leads to a sufficient suppression of the medium opacity for a
blue kilonova to be observed, \ie $X_{\rm Ln} \sim 10^{-5}$. Very similar
values have been obtained in all other BNS models.

Our results, therefore, seem to indicate that a blue-kilonova scenario is
probably unrealistic. As a word of caution, however, we note despite the
three orders of magnitude difference between the expected value and the
one computed here, our conclusions may be biased by an oversimplified
neutrino treatment. A proper radiative-transfer treatment of the
propagation of the radiation in the ejected matter could in fact modify,
at least in part, our results. Indeed, a more sophisticated neutrino
treatment, such as the one employed in Ref. \cite{Foucart2016a}, shows
that can result in a higher values of the electron fraction around the
polar regions. All things considered, our results suggest that while a
blue kilonova component cannot be ruled out conclusively, it also seems
to require an electron-fraction distribution that is considerably
different from the one computed here.

\begin{figure*}
\begin{center}
    \includegraphics[width=2.0\columnwidth]{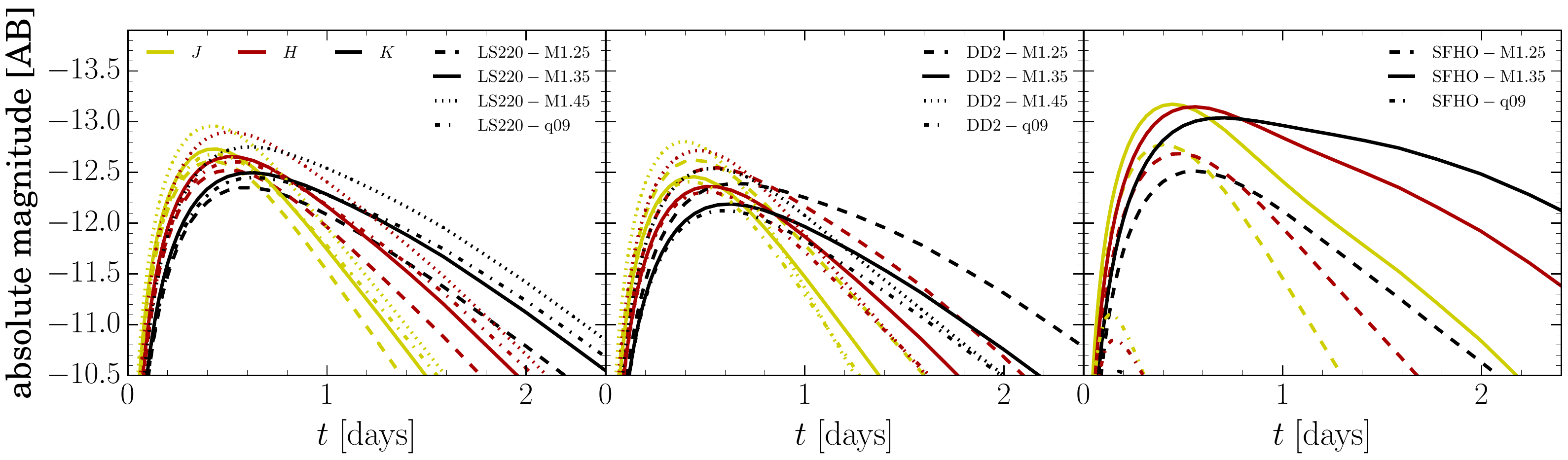}
    \caption{Synthetic light curves in the infrared 2MASS
      $J,H$ and $K$-bands for all of the binaries considered.}
    \label{fig:lightcurves}
    \end{center}
\end{figure*}

\subsection{Kilonova observability}
\label{sec:kilonova_obs}

We assess the observability of the infrared transients associated to the
decay of r-process elements using the simple gray-opacity model of
kilonovae developed in Ref.~\cite{Grossman2014}. The small ejected masses
resulting from our simulations preclude the use of more sophisticated
radiative-transfer treatments (which we leave for a future work) when
these ejecta could be a significant source of opacity (the ``lanthanides
curtain'') for potential secondary outflows, such as magnetically
\cite{Siegel2014} and viscously driven wind from an accretion disk, or
neutrino-driven wind from the hypermassive neutron star
\cite{Perego2014}.

In the model of Ref.~\cite{Grossman2014}, the background dynamical ejecta
are approximated by a homologously expanding spherically symmetric
solution $\rho(r,t) = \rho_0 (t_0/t)^3 (1 - v^2/v_{\rm max}^2)^3$ (also
described in detail in Ref.~\cite{Wollaeger2017}), and $v_{\rm
  max}=2\langle v_\infty\rangle$ from Table~\ref{table:results}. The
luminosity output is computed by integrating the nuclear heating rate
from the nuclear network over the layer of matter from which photons can
diffuse out; a similar model was used also in Refs.~\cite{Perego2014,
  Martin2015, Rosswog2017}. We employ an effective gray opacity
$\kappa=10\ {\rm cm}^2\ {\rm g}^{-1}$, which was recently demonstrated to
reproduce reasonably well the infrared luminosity of lanthanide- and
actinide-contaminated ejecta \cite{Wollaeger2017}. We note that the same
study has shown how the flux in the optical bands is be strongly
suppressed when detailed opacities of lanthanides are used; For this
reason, we consider here only the infrared magnitudes $J$, $H$ and
$K$-bands in the Two Micron All Sky Survey (2MASS) \cite{Skrutskie2006}.

The nuclear heating which powers the kilonova for each model is
calculated with the nuclear network code \texttt{WinNet}
\cite{Winteler2012, Korobkin2012}, using the average electron fraction
$\langle \Ye\rangle$, specific entropy $\langle s\rangle$ and expansion
velocity $\langle v_\infty\rangle$ as given in
Table~\ref{table:results}. We compute the nucleosynthesis yields with
reaction rates based on the finite-range droplet model (FRDM)
\cite{Moeller2012} only. This is motivated by the fact that nuclear mass
models show little discrepancy in the heating rates at epochs around $t
\simeq 1$ day \cite{Rosswog2017}, where the peak magnitudes for our
models are expected.

The resulting peak bolometric luminosities, peak magnitudes in the
infrared bands, and the peak epochs in the $H$-band are presented in
Table~\ref{table:results}, while the light curves in the three infrared
bands (different line colors) are shown in Fig.~\ref{fig:lightcurves},
with different line types referring to the different binaries. 

Clearly, all of our models show a very similar behaviour, peaking around
half a day in the $H$-band and rapidly decreasing in luminosity after one
day, reaching a maximum magnitude of $-13$. We note that these
luminosities are smaller than those normally expected (see, \eg
\cite{Tanaka2016} for a recent review), which peak around magnitude of
$\sim -15$; this difference, however, is not surprising and is mostly due
to the amounts of ejected mass, which is normally assumed to be $\sim
10^{-2}\,M_{\odot}$ and hence at least one order of magnitude larger than
what measured here. With 3-minute $J$-band exposure on the VISTA
telescope \cite{Emerson2004}, these magnitudes result in a detection
horizon of $\sim100$ ${\rm Mpc}$, which, in combination with a very short
time around the peak, makes these light curves extremely difficult to
detect in a follow-up survey. On the other hand, a follow-up observation
of a short GRB with the Hubble Space Telescope could in principle detect
signals of such magnitude up to a redshift $z\sim0.5$, but in such
scenario the kilonova will be most likely outshone by the GRB afterglow.

\section{Constraints on BNS merger rates}
\label{sec:rates}

\begin{figure*}
\begin{center}
\includegraphics[width=1.75\columnwidth]{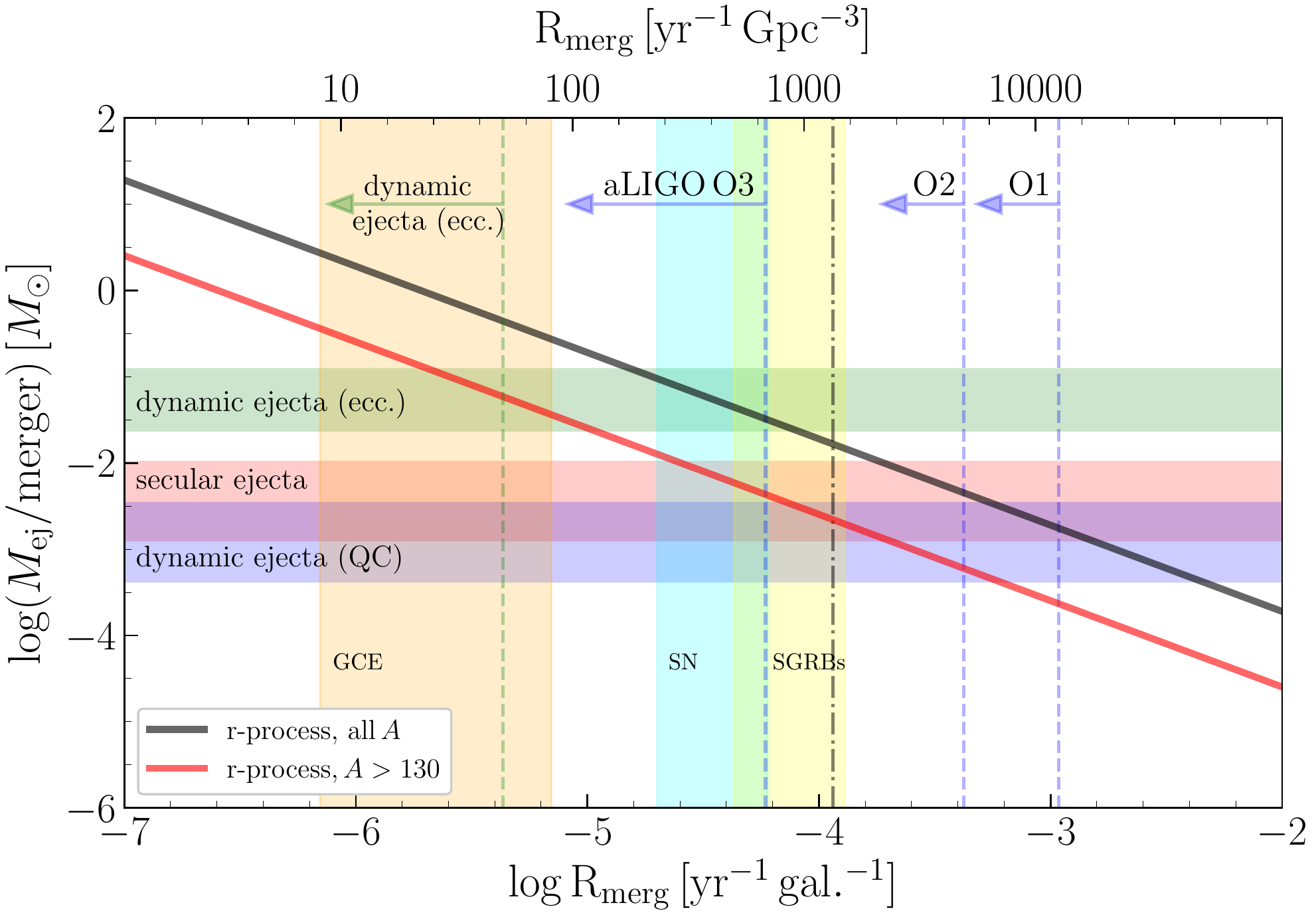}
\caption{Ejected material per merger for a given BNS merger rate
  required to reproduce the observed mass of all (black) and $A>130$
  (red) r-process elements in the Milky Way. The dark blue-shaded regions
  correspond to the range of values of ejected mass reported in
  Table~\ref{table:results}. The red-shaded region corresponds to ejected
  masses from other sources of ejecta. The dashed vertical gray lines
  report the observed, O1, and predicted, O2 and O3, upper bounds on BNS
  mergers from LIGO. The orange, light blue, and yellow shaded regions
  correspond to observational constraints from galactic chemical
  evolution (GCE), supernova (SN), and short gamma-ray bursts (SGRBs)
  population synthesis models as defined in the text. The dot-dashed line
  represents the rate assuming a single NS-NS merger is detected.}
\label{fig:rates}
\end{center}
\end{figure*}

Having assessed the robustness of r-process nucleosynthesis from BNS
mergers and the very good agreement with solar abundances, it still
remains to be established whether the amount of ejected material in a BNS
merger is sufficient to explain the observed amounts of r-process
material in the Milky Way. To this end, and following
Ref.~\cite{Rosswog2017}, we present in Fig.~\ref{fig:rates} the
constraints on the rate of BNS mergers and the required amount of ejected
material needed per merger. More specifically, assuming the total amount
of r-process material in the Galaxy is $M_\mathrm{r,gal} \approx
19,\!000\,\Msol$ and given a certain merger rate -- either per year and
galaxy equivalent ($\mathrm{yr}^{-1}\ \mathrm{gal}^{-1}$, bottom
horizontal axis) or per year and cubic Gigaparsec
($\mathrm{yr}^{-1}\ \mathrm{Gpc}^{-1}$, top horizontal axis) -- the black
line shows the amount of ejected material per merger required to explain
the observed abundances. Similarly, the red line has the same meaning,
but only takes into account elements with $A \gtrsim 130$, with a total
galactic mass of $M_\mathrm{r,gal} \approx 2,\!530 \Msol$
\cite{McMillan2011, Rosswog2017}. The blue-shaded horizontal region
indicates the range of dynamically ejected material from BNS mergers in
quasi-circular (QC) orbits and covered by our simulations as reported in
Table~\ref{table:results}\footnote{The \texttt{SFHO-M1.45} model has been
  omitted because it is not representative.}; the other two shaded horizontal
regions report instead the typical abundances coming from the secular
ejecta (pink-shaded region) or from the dynamic ejecta relative to
mergers of BNSs in eccentric orbits (green-shaded region).

These constraints should be compared with actual measurements of the
merger rates as deduced from different experiments and indicated as
vertical lines. In particular, we show the observed upper bound on BNS
mergers observed in the first LIGO operating run O1, and the predicted
upper bounds for the planned future runs O2 and O3
\cite{Abbott2016h}. Additionally, different population-synthesis models
are also displayed corresponding to galactic chemical evolution (GCE)
\cite{Cote2017}, supernova (SN) \cite{Chruslinska2016}, and SGRBS
\cite{Petrillo2013}.

Overall, both the black and red lines in Fig. \ref{fig:rates} indicate
that the measured ejected masses are sufficient to reproduce the observed
r-process mass abundances given the current O1 reported values for both
all the r-process elements and those with $A \gtrsim 130$. Given the
estimated event rates with the O2 run, the ejected masses are sufficient
to explain $A \gtrsim 130$ elements but not sufficient to explain all
elements. The predicted rates from O3 are rather strict and would require
a lot of ejected material, requiring about an order of magnitude more
than observed in our simulations. 

The constraints placed on the ejecta mass from population synthesis
models are variable and span many orders of magnitude. The constraints
from SGRBS \cite{Petrillo2013} are compatible with the low-mass ejecta
and the upper end of merger rates from SN production models
\cite{Chruslinska2016} can also support low-mass ejecta. In contrast,
galactic chemical evolution models \cite{Cote2017} are more pessimistic
and require significant amounts of ejecta, beyond what has been seen in
current simulations of merging binaries. Additionally, we have added a
predicted merger rate (dash-dot) of a single NS-NS merger detection
assuming a typical LIGO observing run of $9$ months and search area
sensitivity up to $100\Mpc$ \cite{Abbott2016h}. 

As discussed in Sec.~\ref{sec:outflow-prop}, high values of ejected mass
have been reported in some numerical simulations, especially those
employing Newtonian physics. Furthermore, the geodesic criterion that we
employ is a conservative one and only provides a lower bound in the
amount of material ejected; by adopting a different criterion, \eg the
Bernoulli one (see discussion in Appendix \ref{sec:bernoulli}), the
amount of ejected material can increase up to a factor of $4$ for the
same simulation. Should a tension arise between the event rate of BNS
mergers and the ejected material per merger, a possible way out is
offered by mergers from BNSs eccentric orbits. Ejecta masses from these
configurations are in fact much larger \cite{Radice2016,Papenfort2017},
and would be sufficient to explain the observed mass values; however
these events are likely very rare and current constraints are not well
understood \cite{Lee2010}.

Finally, it is important to note that our simulations only focus on the
dynamical ejecta. There are multiple other channels through which
material can be ejected from a merger and they will contribute to the
total amount of r-process elements created. For example, simulations of
neutrino-driven winds have found similar amounts of ejected material as
dynamical ejecta \cite{Fujibayashi2017}. Likewise, matter ejected from a
BH-torus system could be as high as $0.1\,\Msol$, as estimated
semi-analytically in Ref. \cite{Giacomazzo2012b}.  This suggests that
even if the mass ejected from a single channel is alone insufficient to
explain the observed r-process masses, the combination of all ejected
material from a BNS merger is likely to. In this sense, the blue-shaded
horizontal region only represents a lower bound on the total ejected
material.

\section{Conclusions}
\label{sec:CONCLUSIONS}

The material ejected dynamically from the merger of neutron stars
binaries is neutron-rich and its nucleosynthesis could provide the
astrophysical site for the production of heavy elements in the universe,
together with the unambiguous confirmation that the merger of neutron
stars is behind the origin of SGRBs.

Making use of fully general-relativistic calculations of the inspiral and
merger of binary systems of neutron stars, we have investigated the role
of initial masses, mass ratio, and EOS on the r-process nucleosynthesis
taking place in the dynamical ejecta from BNS mergers. To do so, we have
made use of tracer particles that allow us to follow the fluid and that
can be used to extrapolate the fluid properties to the late times needed
to run nuclear networks codes, together with a simplified neutrino
leakage scheme. 

Among the several results reported, three deserve special mention. First,
we have shown that the r-process nucleosynthesis from BNS mergers is very
robust and essentially universal in that it depends only very weakly on
the properties of the binary system, such as the EOS, the total mass or
the mass. In all cases considered, in fact, and modulo small differences,
the yields of our nuclear-reaction networks are in very good agreement
with the solar abundances for mass numbers $A \gtrsim 120$. While similar
conclusions have been reported before, the confirmation coming from our
study strengthens the evidence that BNS mergers are the site of
production of the r-process elements in the galaxy.

Second, we have employed two different approaches to measure the amount
of matter ejected dynamically and found that it is $\lesssim
10^{-3}\,M_{\odot}$, which is smaller than what usually assumed. There
are a number of factors that need to be taken into account when deriving
these estimates, namely: the EOS, the neutrino treatment, the criterion
for unboundness, the resolution, the numerical methods used. Although
these systematic factors can lead to differences as large as one order of
magnitude even for the same initial data, we find it unlikely that the
mass ejected dynamically can ever reach the values sometimes assumed in
the literature of $10^{-2}-10^{-1}\,M_{\odot}$. Clearly, a more detailed
and comparative study is necessary to better constrain the uncertainties
behind the amount of mass lost by these systems.

Third, using a simplified and gray-opacity model we have assessed the
observability of the infrared transients associated to the decay of
r-process elements, \ie of the kilonova emission. We have found that all
of our binaries show a very similar behaviour, peaking around $\sim 1/2$
day in the $H$-band and rapidly decreasing in luminosity after one day,
reaching a maximum magnitude of $-13$. These rather low luminosities are
most probably the direct consequence of the small amounts of ejected
matter, thus making the prospects for detecting kilonovae rather
limited. More sophisticated calculations with improved neutrino
treatments will be needed to settle this issue conclusively.

As a final remark we note that even though the r-process abundance
pattern does not give us simple clues to the original BNS parameters, \eg
it does not allow to disentangle various EOS and mass configurations,
there are distinguishing features in the ensuing kilonova signal
relatable through the difference in ejecta properties obtained in our
simulations. In particular, we have found that softer EOSs tend to result
in a higher average electron fractions, which implies differences in the
type of kilonova produced (blue vs red kilonovae). Additionally, we have
found that this difference in electron fraction is highly angular
dependent with higher electron fractions around the polar regions and
lower along the orbital plane. Even though there is significantly less
material ejected along the poles versus the plane, our simulations show
that the simplified models of kilonova modelling, such as that of a
homogeneously expanding group of material, need to be adjusted to account
for this anisotropic emission. We reserve the investigation of this issue
to future studies, where an improved neutrino treatment will also be
implemented.


\section*{Acknowledgements}

It is a pleasure to thank M. Eichler, M. Hempel, E. Most, and D. Radice
for useful discussions. LB thanks the attendees of the MICRA 2017
Conference at Michigan State University and of the INT Workshop 17-2b
``Electromagnetic Signatures of r-process Nucleosynthesis in Neutron Star
Binary Mergers'' for useful discussions related to this work. This
research is supported in part by the ERC Synergy Grant ``BlackHoleCam:
Imaging the Event Horizon of Black Holes" (Grant No. 610058), the ERC
starting grant ``EUROPIUM'' (Grant No. 677912), by ``NewCompStar'', COST
Action MP1304, by the LOEWE-Program in the Helmholtz International Center
(HIC) for FAIR, by the European Union's Horizon 2020 Research and
Innovation Programme (Grant 671698) (call FETHPC-1-2014, project
ExaHyPE), by the Helmholtz-University Young Investigator Grant
No. VH-NG-825, and by the BMBF under Grant No. 05P15RDFN1. LB and FG are
supported by HIC for FAIR and the graduate school HGS-HIRe. The
simulations were performed on the SuperMUC cluster at the LRZ in
Garching, on the LOEWE cluster in CSC in Frankfurt, on the HazelHen
cluster at the HLRS in Stuttgart.

\appendix

\section{Comparison of criteria for unbound material}
\label{sec:bernoulli}

\begin{figure}[t]
\begin{center}
    \includegraphics[width=\columnwidth]{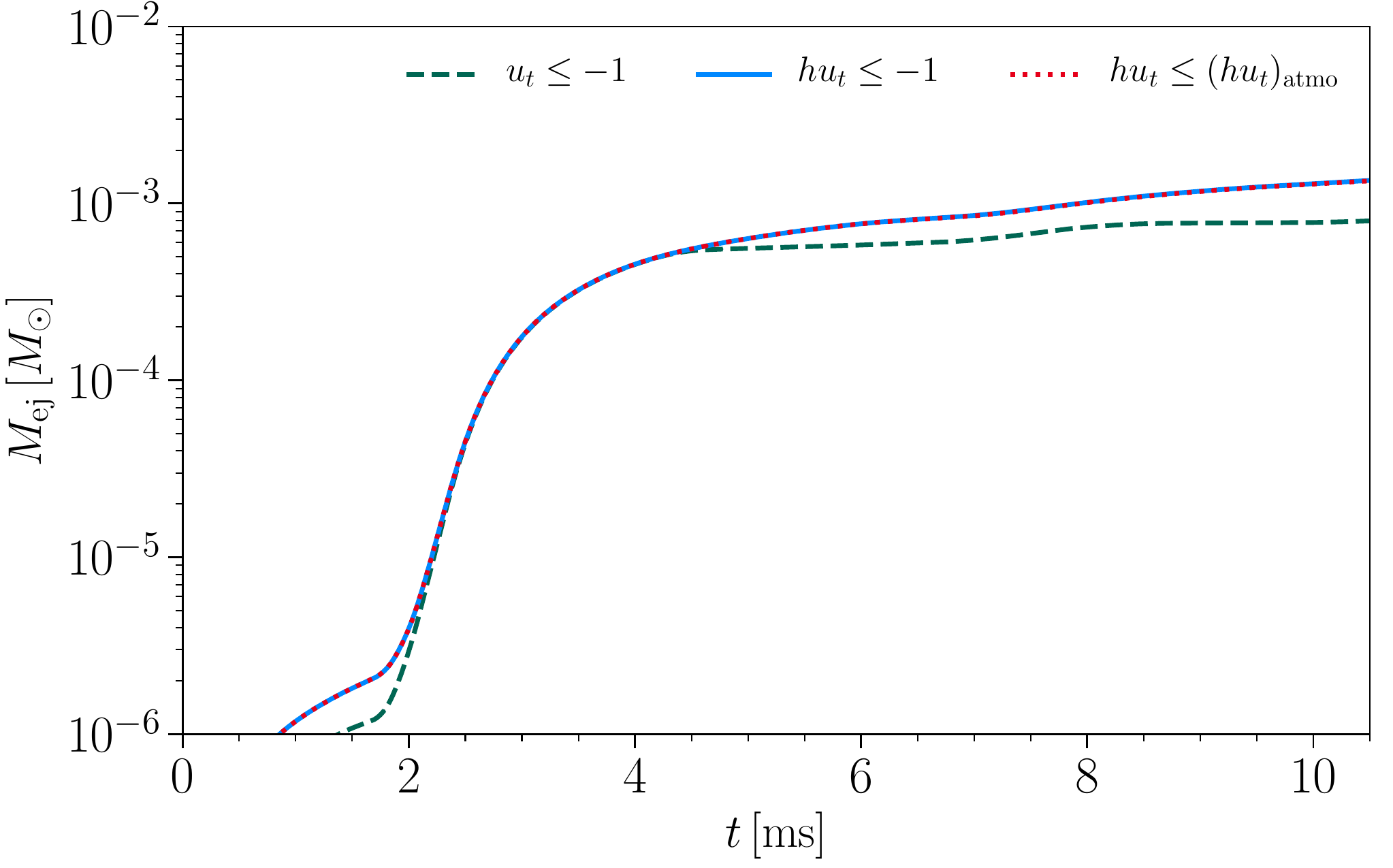}
    \caption{Mass ejection according to different unboundness criteria
      for the \texttt{LS220-M1.35} model. In green is the geodesic
      criterion, blue is the original Bernoulli one, and red is the
      modified Bernoulli thresholded on the atmosphere value. All values
      have been measured through a detector at $300\km$.}
    \label{fig:mass-comp}
\end{center}
\end{figure}

In Sec.~\ref{sec:unboundcrit}, we introduced the criteria by which we
determine unbound material. In Secs.
\ref{sec:overview}--\ref{sec:velocity}, we considered only the geodesic
criterion for determining unbound material. The justification for this
choice of the geodesic criterion is its simplicity and the fact that it
provides a lower bound for the total ejected material
\citep{Kastaun2014}.  An additional benefit of the geodesic criterion is
that it does not implicitly depend on the EOS selected, while the
Bernoulli criterion, through the enthalpy $h$, does. This implies that a
fluid element, with the same rest-mass density, temperature, and electron
fraction, can be unbound or bound depending on the EOS through the
Bernoulli criterion.  This is a relatively minor trade-off since through
the introduction of the enthalpy, the effects of pressure and temperature
are taken into account.

\begin{figure*}[t]
\begin{center}
\includegraphics[width=2.0\columnwidth]{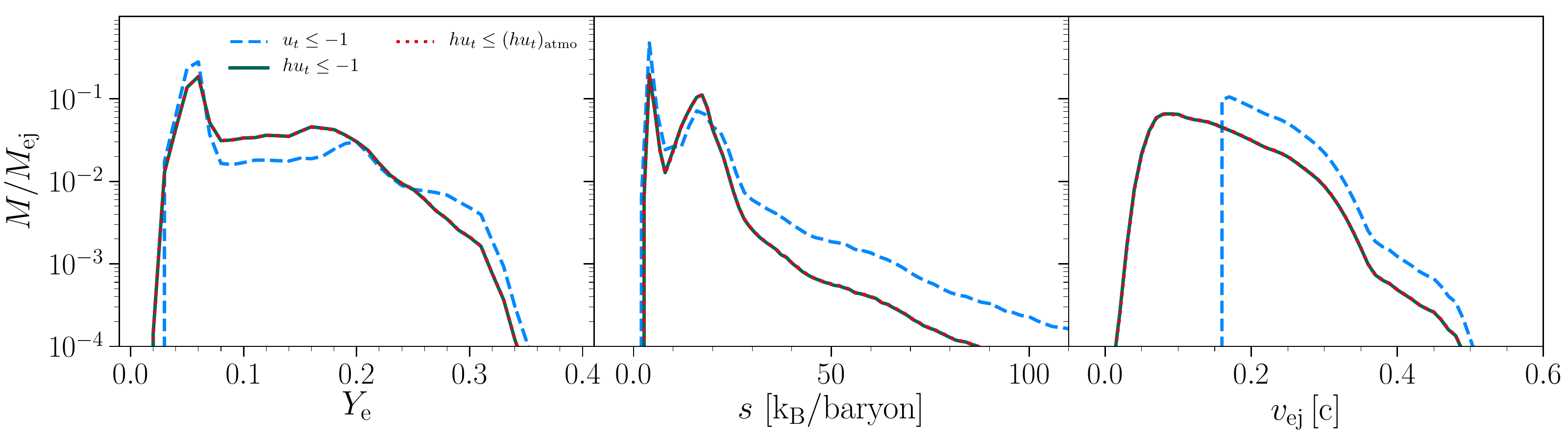}
\caption{Comparison of the mass distribution of electron fraction,
    specific entropy and ejecta velocity in the ejected matter of the
    representative model \texttt{LS220-M1.35} for the three unboundedness
    criteria.}
\label{fig:bern-distro}
\end{center}
\end{figure*}

Since the specific enthalpy is always grater than one, $h\ge 1$, we have
that
\begin{equation}
    |h\,u_{t}| \ge |u_{t}|\,,
\end{equation}
and thus the Bernoulli criterion will always result in more material
becoming unbound. However, a slight modification of this formula is
required. In our simulations, we have an atmosphere that acts as a lower
bound for the hydrodynamical quantities. As discussed in Sec.
\ref{sec:mass-eject} we have chosen to evaluate the ejecta at $300\km$
away from the merger remnant to avoid atmospheric effects. But due to the
introduction of the enthalpy, we need to ensure that we are sufficiently
above the atmosphere to avoid unphysical atmosphere values entering our
calculations. To achieve this, instead of defining unbound elements as
satisfying the relation $hu_{t}\le -1$, we consider the following
modified criterion
\begin{equation}
h\,u_{t} \le (h\,u_{t})|_{\rm atmo}\,,
\end{equation}
where we evaluate the $hu_{t}$ at the values set by our atmosphere setup,
which is EOS-dependent. For example, for the LS220 EOS this
term assumes the value
\begin{equation}
    h\,u_{t} \le -1.000163\,,
\end{equation}
instead of $-1$. Even though this difference is small, the modified
constraint does exclude some material from being considered as ejected.

We proceed at comparing the results of the geodesic, the original and the
modified Bernoulli criteria in the fiducial case of the
\texttt{LS220-M1.35} model. In Fig. \ref{fig:mass-comp}, we show the
differences between the three selection criteria in the mass ejection
curve. Overall, the behaviour for the different criteria is similar, with
an ejection phase beginning approximately $2\,\ms$ after merger followed
by a decrease in the amount of ejected mass. While the geodesic-selected
material approaches a constant value, both Bernoulli criteria show a
slightly longer increasing phase before settling to a constant. In Table
\ref{table:results-bern}, we show the comparison of the ejected material
for the three criteria and find that by selecting one of the Bernoulli
criteria, we obtain approximately $2.5$ times as much ejected material
when compared to the geodesic one. This increase in the amount of ejecta
is similar across all simulations we have performed: the ejected mass is
larger by a factor $1.5$ to $4$ with the Bernoulli criterion as compared
with the geodesic one.

\begin{table}[!h]
  \begin{center}
    \begin{tabular}{|r||cccc|}
      \hline
      &  $M_{\rm{ej}}$ & $\langle \Ye \rangle $ & $\langle s
      \rangle $ & $\langle v_{\rm{ej}} \rangle$  \\
      Criterion & $[10^{-3}\, \Msol]$ & -  & $[\kB]$ & $[10^{-1} \rm{c}]$ \\
      \hline
      \texttt{geodesic} & 0.82 & 0.10 &12.3 &2.2\\
      \texttt{Bernoulli} & 2.09 & 0.11 &13.8 &1.5\\
      \texttt{modified Bernoulli} & 2.07 & 0.11 &13.1 &1.5\\
      \hline
    \end{tabular}
    \caption{Average values of the ejected mass, electron fraction,
      specific entropy and ejecta velocity for different unboundedness
      criteria in the representative \texttt{LS220-M1.35} model.}
    \label{table:results-bern}
  \end{center}
\end{table}

In Fig. \ref{fig:bern-distro}, we plot again the mass distribution in the
ejecta of the various quantities relevant for r-process nucleosynthesis,
again for the representative \texttt{LS220-M1.35} model.  Additionally,
the average values are summarized in Table \ref{table:results-bern}. For
the electron fraction and entropy, we do not see drastic changes and the
overall structure of the distribution between different criteria. In both
cases, there is a slight increase in entropy and $\Ye$ which is to be
expected. With both Bernoulli criteria, taking the enthalpy into account
includes some thermodynamic effects which will result in more material
being ejected due to shock heating.  This implies a higher entropy and
additionally, more material to undergo neutrino interactions. However,
the effects are minimal and the overall nucleosynthesis process will be
essentially unaffected.

Finally, the most striking difference is in the $v_{\rm ej}$ quantity. In
the geodesic criterion, interpreted in its Newtonian limit, a fluid
element has to have non-zero velocity at infinity to be considered
unbound. This implies a lower cutoff in the velocity distribution, as
slowly moving elements, even though able to cross a given detector
surface, would not be considered unbound. For both Bernoulli criteria
this strict requirement is relaxed by the presence of the enthalpy, which
acts as a multiplicative factor larger than one. This means that even
slowly moving elements, provided they have sufficiently high enthalpy,
would be counted as unbound, and so the velocity distribution acquires a
lower end tail and its mean is shifted towards lower values.

\section{Additional information on ejecta morphology}
\label{sec:add_morphology}

\begin{figure*}[!t]
    \centering
    \includegraphics[width=2.0\columnwidth]{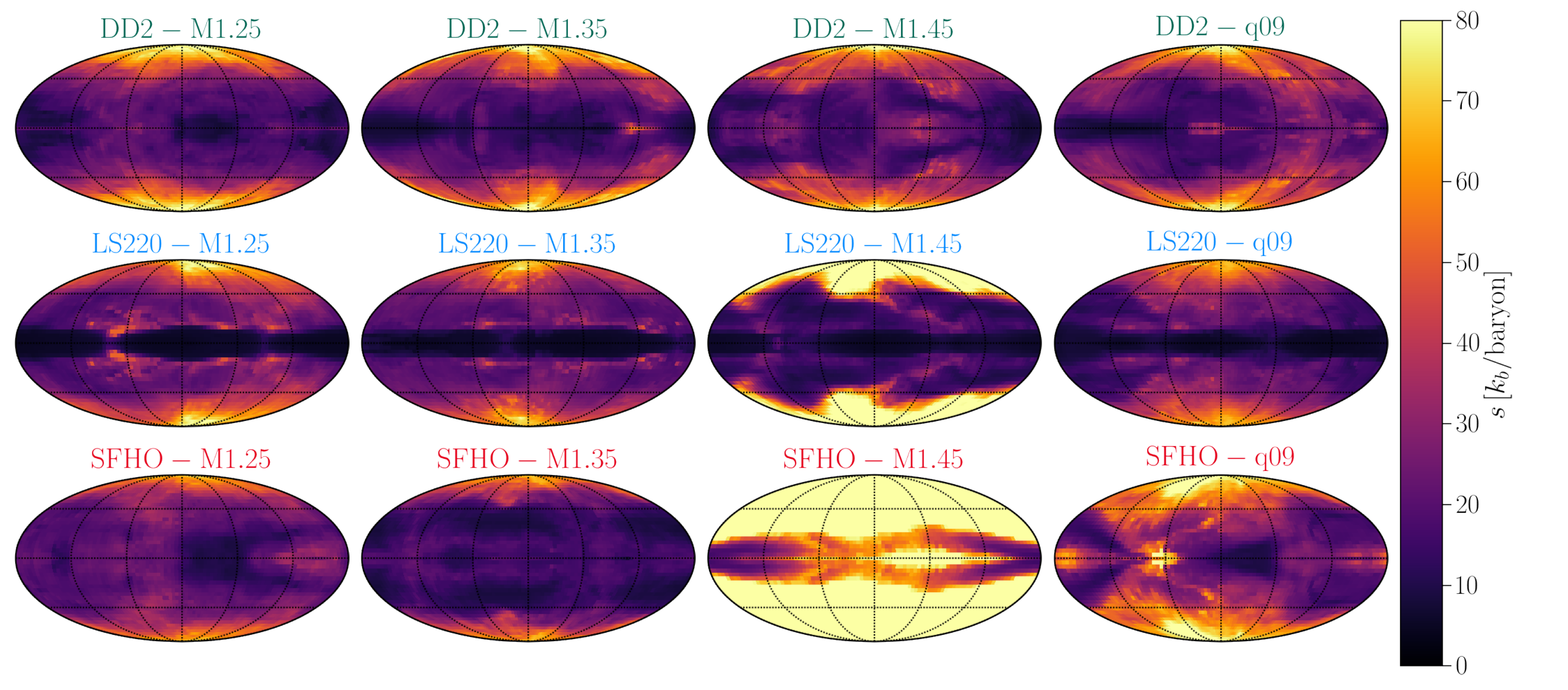}
    \caption{The same as in Fig. \ref{fig:mass2D} but for the specific
      entropy averaged over the ejected mass.}
    \label{fig:s_bar2D}
\end{figure*}

In Sec. \ref{sec:morphology} we analyzed the morphology of the ejected
matter, focusing on the angular distribution of the ejected mass and
electron fraction. In this Appendix we report the results of the
analogous analysis performed on the specific entropy and ejecta velocity
distribution.

In the case of the specific entropy, similar observations hold true as
for the electron fraction distribution, see Fig.~\ref{fig:s_bar2D}: the
entropy anticorrelates with the ejected matter. Regions close to the
orbital plane tend to have specific entropy values of $10\,\kB/{\rm
baryon}$ or lower. This is easily seen in the case of the
LS220 EOS simulations.

Around the poles values of the entropy can be very high. These
corresponds to the tails shown in Fig.  \ref{fig:s-distro}, extending to
specific entropies of $200\,\kB$ and above. The angular size of the polar
high-entropy regions depends on the EOS and mass configuration if each
run, but also in this case it appears to (anti-) correlate with the
angular distribution of the mass ejection: in cases in which the ejection
is strongly focused on the orbital plane, \eg model \texttt{LS220-M1.25},
higher values of the specific entropy at lower latitudes can be reached,
and vice versa.

The tendency of regions with low ejected mass to show higher entropy is
easily understood in terms of shock-heating in rarefied environments: the
efficiency of shock-heating is enhanced in low-density media, where less
thermal energy is required to heat the material to higher temperatures
and raise the entropy. In this sense, the case of the \texttt{SFHO-M1.45}
model is particularly striking: in this simulation most of the ejected
material is at extremely high specific entropy. As observed in the
previous discussion, this model also ejects an almost negligible amount
of mass, greatly enhancing the shock-heating efficiency.

\begin{figure*}[!t]
    \centering
    \includegraphics[width=2.0\columnwidth]{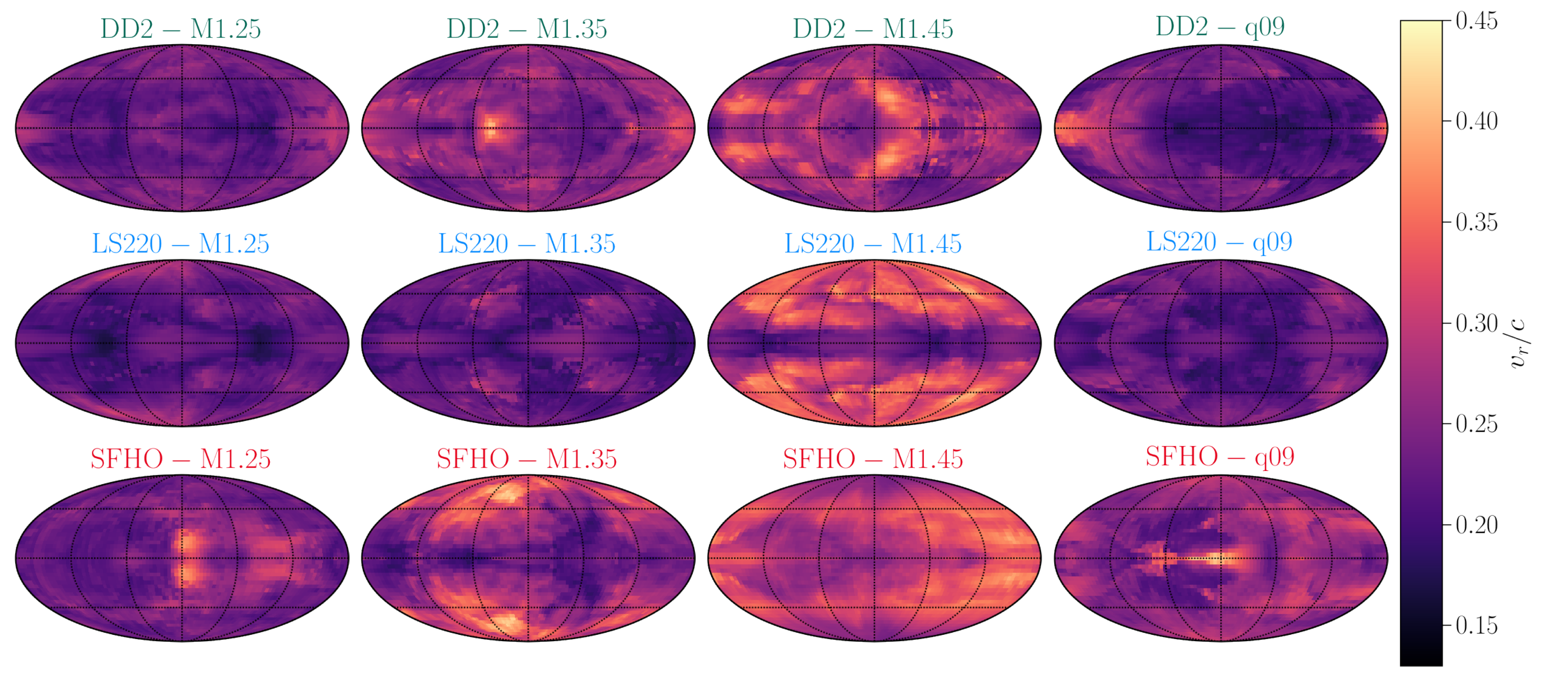}
    \caption{The same as in Fig. \ref{fig:mass2D} but for the ejecta
      velocity averaged over the ejected mass.}
    \label{fig:v_bar2D}
\end{figure*}

The velocity distribution, shown in Fig.~\ref{fig:v_bar2D}, is instead
rather peculiar. For many models, especially the lower mass ones,
including the unequal-mass models in the rightmost column of the figure,
the material appears to be expanding at the same velocity in most
directions, save for a few ``hot'' or ``'cold'' spots of limited angular
size. In the three higher mass models, shown in the third column of
Fig.~\ref{fig:v_bar2D}, some large-scale structure could be present, but
there is no evidence of the correlation observed for the electron
fraction or entropy.

Note that, as mentioned in Sec. \ref{sec:morphology}, the angular
distribution of the hydrodynamical and thermodynamical properties of the
ejected matter is computed disregarding bound fluid elements (\cf Eq.
\eqref{eq:morph}), \ie fluid elements that do not satisfy one of the
criteria for unboundedness outlined in Sec. \ref{sec:unboundcrit} and
Appendix \ref{sec:bernoulli}. In particular, Figs.  \ref{fig:mass2D},
\ref{fig:Y_e_bar2D}, \ref{fig:s_bar2D} and \ref{fig:v_bar2D} have been
obtained by considering as unbound fluid elements satisfying the geodesic
criterion. If the Bernoulli criterion is employed, the morphology of the
ejecta is qualitatively unchanged. This is particularly the case when
examining the angular dependence of the ejected mass, electron fraction
and specific entropy. The most prominent difference is that the mass
ejection extends to higher latitudes, instead of being mostly confined to
the orbital plane as in the case of the Geodesic criterion. The electron
fraction and specific entropy, being anticorrelated with the ejected
mass, follow a similar distribution, where however the regions of high
$\Ye$ and $\s$ close to the poles show a reduced extent. This is to be
expected, since the Bernoulli criterion is less restrictive than the
geodesic one in defining fluid elements as unbound.

The most striking difference in the ejecta morphology due to the
unboundedness criterion is in the distribution of the ejecta velocity. As
shown in Appendix \ref{sec:bernoulli}, \cf Fig. \ref{fig:bern-distro},
the distribution of the ejected mass with respect to its velocity is
extends to very low velocity values when the Bernoulli criterion is
considered. This is reflected in the angular distribution of the
velocity, which for all models where significant mass ejection takes
place, has an average value on all angular directions of $\sim0.15\,c$,
significantly lower than for the Geodesic criterion. The
Bernoulli-computed distribution of the velocity is also less anisotropic
than the geodesic one.

\bibliographystyle{apsrev4-1}
%


\end{document}